\documentclass[%
 superscriptaddress,twocolumn,
 amsmath,amssymb,
aps,
prx,
]{revtex4-1}

\usepackage{graphicx}
\usepackage{dcolumn}
\usepackage{bm}
\usepackage{lipsum,graphicx}
\usepackage[export]{adjustbox}
\usepackage{float}
\usepackage[colorlinks,allcolors=blue]{hyperref}
\usepackage{physics}

\usepackage{todonotes}

\graphicspath{{figures/}}

\begin{document}

\title{Topological superconductivity in nanowires proximate to a diffusive superconductor-magnetic insulator bilayer}

\author{Aleksei Khindanov}
\email[]{khindanov@ucsb.edu}
\affiliation{
	Department of Physics, University of California, Santa Barbara, California 93106, USA
}

\author{Jason Alicea}
\affiliation{Department of Physics and Institute for Quantum Information and Matter,
California Institute of Technology, Pasadena, California 91125 USA}
\affiliation{Walter Burke Institute for Theoretical Physics, California Institute of Technology, Pasadena, California 91125 USA}

\author{Patrick Lee}
\affiliation{Department of Physics, Massachusetts Institute of Technology, Cambridge, Massachusetts 02139, USA}

\author{William S. Cole}
\affiliation{Microsoft Quantum, Station Q, Santa Barbara, California
	93106, USA}

\author{Andrey E. Antipov}
\affiliation{Microsoft Quantum, Station Q, Santa Barbara, California
	93106, USA}

\date{\today}

\begin{abstract}

We study semiconductor nanowires coupled to a bilayer of a disordered superconductor and a magnetic insulator, motivated by recent experiments reporting possible Majorana-zero-mode signatures in related architectures.
Specifically, we pursue a quasiclassical Usadel equation approach that treats superconductivity in the bilayer self-consistently in the presence of spin-orbit scattering, magnetic-impurity scattering, and Zeeman splitting induced by both the magnetic insulator and a supplemental applied field.
Within this framework we explore prospects for engineering topological superconductivity in a nanowire proximate to the bilayer.
We find that a magnetic-insulator-induced Zeeman splitting, mediated through the superconductor alone, cannot induce a topological phase since the destruction of superconductivity (i.e., Clogston limit) preempts the required regime in which the nanowire’s Zeeman energy exceeds the induced pairing strength.
However, this Zeeman splitting does reduce the critical applied field needed to access the topological phase transition, with fields antiparallel to the magnetization of the magnetic insulator having an optimal effect.
Finally, we show that magnetic-impurity scattering degrades the topological phase, and spin-orbit scattering, if present in the superconductor, pushes the Clogston limit to higher fields yet simultaneously increases the critical applied field strength.

\end{abstract}

\maketitle

\section{\label{sec:Intro}Introduction}

Spatially separated ``Majorana'' zero-energy modes in topological superconductors encode an unusually robust ground state degeneracy through the presence or absence of quasiparticle fermionic excitations shared nonlocally by each Majorana pair.
These represent an appealing candidate for quantum information storage and processing that is passively robust to local perturbations; i.e., a topological quantum computer~\cite{Freedman2003, Nayak2008, Dassarma2015}.
For a variety of reasons --- the paucity of intrinsic topological materials, the maturity and scalability of semiconductor technology --- much of the effort to date has been toward the realization of Majoranas in hybrid systems of relatively conventional components~\cite{Alicea2010, Lutchyn2010, Oreg2010}: a narrow gap semiconductor with strong spin-orbit coupling and large $g$-factor, proximitized by a thin $s$-wave superconducting film and subjected to a magnetic field parallel to the film.
Each ingredient in this recipe is crucial, and combining them is not a trivial task \cite{Lutchyn2018}.
For example, the ``topological gap" to quasiparticle excitations outside the degenerate ground-state space is bounded from above by the proximity-induced gap at zero field, and yet too-strong coupling between the semiconductor and superconductor (to maximize this gap) results in an unwanted decrease of the effective $g$-factor~\cite{Antipov2018} while exposing the subgap states to disorder in the superconductor~\cite{Lutchyn2012, Cole2016}.
The magnetic field is needed to open a gap between helicity bands of the semiconductor, stabilizing effective $p$-wave pairing, but competes with superconductivity, while the orbital effect of the field in the semiconductor is also generally antagonistic to a robust topological phase~\cite{Nijholt2016, Winkler2019}.
The sensitivity to field alignment poses restrictions to architectures based on networks of wires~\cite{Karzig2017}.

\begin{figure}[t]
	\includegraphics[width = \columnwidth]{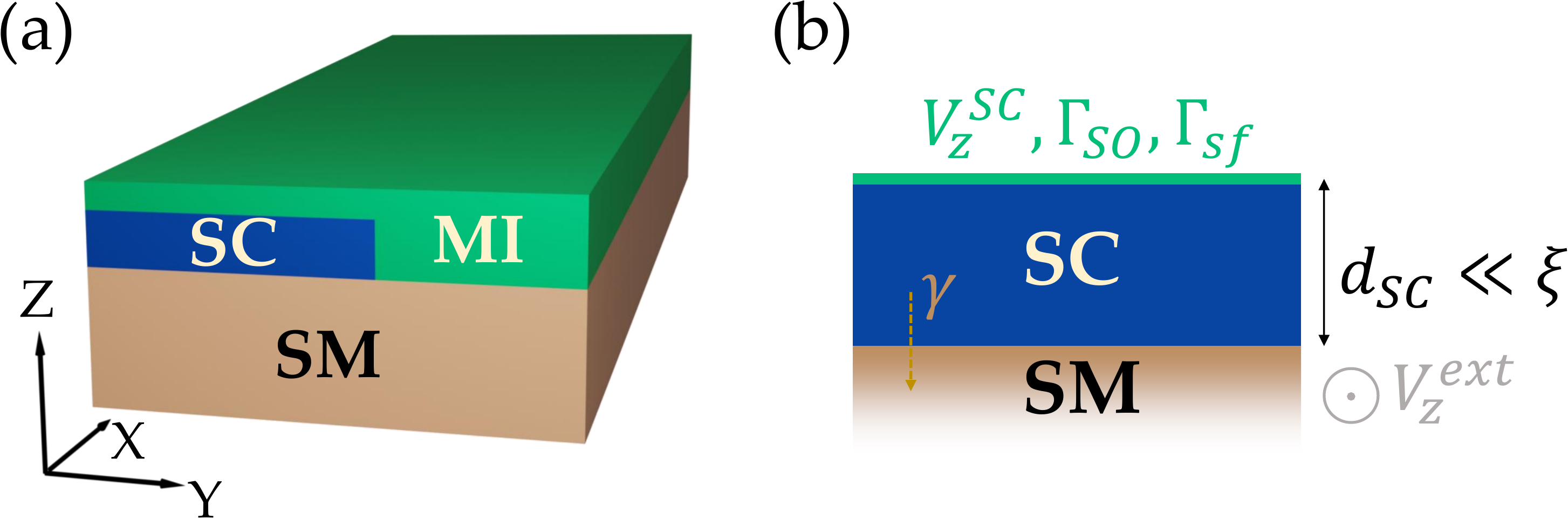}
	\caption{(a) Semiconductor (SM) nanowire proximitized by a superconductor-magnetic insulator (SC-MI) bilayer. (b) Effects of the magnetic insulator on the superconductor can be described by an appropriate boundary condition marked with green.}
	\label{fig:Setup}
\end{figure}

Thus, material optimization continues to play a critical role going forward. One can improve on the material composition by optimizing the semiconductor or the superconductor (as well as their interface) or by eliminating the magnetic field.
While the quality of the semiconductor continues to receive significant attention, even the best possible devices remain limited by the gap in the parent superconductor and the restrictions imposed by the applied magnetic field.
Optimizing the superconductor or eliminating the external field are therefore promising paths to future breakthroughs.

Recently the first attempts toward zero-external-field topological superconductivity have been made ~\cite{Vaitiekenas2020, Liu2020a}.
A typical setup of a heterostructure consists of semiconducting (SM), superconducting (SC) and magnetic insulator (MI) parts connected together as shown schematically in Fig.~\ref{fig:Setup}(a). Here the magnetic insulator such as EuS, induces a Zeeman spin splitting by virtual tunneling.
Previous experiments on SC-MI heterostructures observed a proximity-induced spin splitting of the superconducting density of states~\cite{Tedrow1986,Hao1991,Wolf2014,Strambini2017,Hijano2021}.
In Ref.~\onlinecite{Liu2020} a defect-free heterostructure between an InAs semiconductor wire and EuS has been prepared and studied.
Negligible direct magnetization of InAs was reported.
The authors of Refs.~\onlinecite{Vaitiekenas2020, Liu2020a} further coupled EuS to Al and InAs with the ultimate goal of inducing zero-field topological superconductivity.
Epitaxial growth on different wire facets was achieved and the stray magnetic field generated by EuS was found to be insufficient for inducing a topological phase~\cite{Liu2020a}.
Furthermore, conductance spectroscopy revealed a distinctly different behavior for different geometries of the system~\cite{Vaitiekenas2020}.
The spectroscopic features of wires with EuS and Al residing on different facets of the InAs nanowire were found to be similar to those of nanowires without EuS.
However in geometries with overlapping EuS and Al, the authors of Ref.~\onlinecite{Vaitiekenas2020} observed zero-bias peaks in the differential conductance.

These experiments have inspired several related theoretical investigations~\cite{Woods2020,Liu2020b,Escribano2020,Maiani2020,Poyhonen2020,Langbehn2020}.
For basic reasons~\cite{Poyhonen2020} it is impossible to achieve topological superconductivity solely by proximity to a spin-split conventional superconductor.
Thus the authors of Refs.~\cite{Woods2020,Liu2020b,Escribano2020} rely on spin-splitting in the semiconductor arising from the EuS, directly, as in the original MI-based proposal~\cite{Sau2010}, while the authors of Ref.~\cite{Maiani2020} suggest using EuS as a spin-filter barrier between the superconductor and the semiconductor.
Given the large overlap between the MI and SC, the interaction between the superconductor and the ferromagnet is likely to play an important role.
The authors of Ref.~\cite{Langbehn2020} considered proximity effects self-consistently in different stack geometries and studied potential topological phases.
The disorder-free approximation introduced several important caveats, such as the excess stability of the superconductor to the exchange field induced by the MI, i.e., lack of Clogston limit with instead a second-order transition to the normal phase at sufficiently high exchange field~\cite{Tokuyasu1988}.

It is important to note that the superconductors that have been used in proximity heterostructure experiments, such as Al, NbTiN, Sn, and Pb are all diffusive either due to intrinsic disorder or oxidation on the surface.
On the other hand, the optimal platform for Majorana nanowires should feature ``clean'' semiconductors with defects minimized \cite{Sau2012}.
Therefore one needs to put forward a theoretical framework that includes self-consistent superconducting effects in different parts of the system, disorder scattering, as well as proximity effects.
One such approach would be a self-consistent microscopic Bogoliubov-de Gennes treatment.
However, in practice the need to include phenomena at disparate lengthscales --- ranging between angstroms to hundreds of nanometers and governed by the Fermi wavelengths, superconducting and magnetic coherence lengths as well as disorder scattering mean free path --- makes it prohibitive for numerical real-space calculations, thus necessitating an effective theory.
An established approach for self-consistent superconducting calculations is the quasiclassical framework, which utilizes the relative smallness of the Fermi wavelength compared to characteristic length scales of the system, allowing one to focus the theoretical description in the narrow range of energies close to the Fermi surface.
Such an approach has been utilized before for one-dimensional models of nanowires, including the single-band clean and disordered cases, with real-space profiles of wavefunctions established \cite{Stanev2014} and stability to disorder calculated \cite{Hui2014}.
The method has been extended to the multi-subband regime \cite{Neven2013} and two-dimensional superconductors \cite{Lu2020}.
However, self-consistent effects in the superconductor have not been considered.

In this paper we focus on the physics of MI/SC/SM stack. Instead of directly modeling the geometry shown in Fig.~\ref{fig:Setup}(a), we consider a variety of scattering mechanisms at the interface between the MI and the SC, as shown in Fig.~\ref{fig:Setup}(b). We assume that the effect of the interface between the MI and the SM, shown on the right side of Fig. ~\ref{fig:Setup}(a), can be modeled by an effective bulk Zeeman field in the SM. 
To this end we develop a two-step approach. First, utilizing the short mean free paths in superconductors, we calculate the properties of the MI-SC bilayer using the Usadel equation~\cite{Usadel1970}, and, second, we use this result as a boundary condition for the nanowire model.
We apply the Usadel equation to compute the pair potential, critical temperature, and the density of states in the superconductor.
Various physical processes such as applied magnetic field, exchange field from the magnetic insulator and scattering off magnetic and/or spin-orbit impurities in the SC are incorporated.
The superconducting proximity effect is then readily described by the solution of the Usadel equation.

The rest of the paper is organized as follows. 
Section \ref{sec:methods} describes a theoretical framework we utilize throughout the paper.
Specifically, in Section~\ref{sec:Usadel} we focus on self-consistent superconductivity in the parent SC and review the Usadel equation. In Section~\ref{sec:sc-sm_proximity} we describe the superconductor-semiconductor proximity effect and introduce framework for computing the properties of the topological phase.
The main results of our work are demonstrated in Section~\ref{sec:results}: First, using the Usadel equation we calculate the density of states, pair potential and critical temperature of the SC as a function of Zeeman energy and spin-orbit/magnetic scattering. 
We demonstrate the destruction of the superconducting phase by large Zeeman field and/or magnetic scattering and quenching of the Zeeman effect by the intrinsic spin-orbit scattering.
Next, with the help of the obtained values of the pair potential, we study SC-SM proximity effect and infer conditions for the topological phase transition in the heterostructure. 
We find that an additional Zeeman field needs to be added to the semiconductor to induce the topological phase and we provide an analytical estimate for this field.
Then, we compute the topological gap and study its dependence on external magnetic field when Zeeman splitting and/or spin-orbit and magnetic scattering is present in the SC.
We will be focused on the experimentally relevant regime of a thin superconductor compared to its coherence length (see Fig.~\ref{fig:Setup}b).
Furthermore, we are interested in the regimes of small applied external magnetic fields, thus in this study we ignore the orbital contribution of the applied field.
We show that only one orientation of magnetic field is preferable for the existence of the topological phase.
Magnetic scattering in general is always detrimental to the topological phase, whereas intrinsic spin-orbit scattering in the superconductor helps in sustaining magnetic fields, but increases the critical field required for the topological phase.
Implications of our work for engineering topological systems with SC-MI bilayers and more complicated stacks are given in concluding remarks in Section~\ref{sec:conclusion}.

\section{\label{sec:methods}Method}

\subsection{\label{sec:Usadel}Usadel equation}

The Usadel equation is a nonlinear second-order differential equation for the quasiclassical Green's function of a superconductor.
Although it is a standard method for describing superconductors \cite{Ivanov2006}, for completeness of the presentation we introduce it in this section. The detailed derivation can be found in Ref.~\cite{Bergeret2005}.
The Usadel equation is valid in the limit $\lambda_F \ll l_{MFP} \ll \xi$, where $\lambda_F$ is the metallic Fermi velocity, $l_{MFP}$ is the mean free path and $\xi$ is the superconducting coherence length.
For typical $s$-wave superconductors used in Majorana nanowires, such as Al, this approximation holds, as $\lambda_F \simeq 1 \AA$, $l_{MFP} \simeq 20 \mathrm{nm}$ \cite{Gall2016}, and $\xi \simeq 300 \mathrm{nm}$ \cite{Romijn1982}.

The starting point is the Gor'kov equation for the superconducting Green's function $\check G_{SC}(i\omega_n, \boldsymbol{r_1},\boldsymbol{r_2})$ \cite{Gorkov1958} describing an excitation in Nambu space between spatial coordinates $r_1$ and $r_2$ at the (imaginary) frequency $i\omega_n$.
We will use the mixed real- and momentum- space representation $\check G_{SC}(\boldsymbol{r},\boldsymbol{k})$ obtained by the Wigner transform to the center-of-mass coordinates $\boldsymbol{r}\equiv(\boldsymbol{r_1}+\boldsymbol{r_2})/2$ and a Fourier transform over the relative coordinate $\boldsymbol{r_1}-\boldsymbol{r_2}\to \boldsymbol{k}$. Taking advantage of the short Fermi wavelength in the superconductor, one can apply the quasiclassical approximation and integrate out the magnitude of the relative momenta on the Fermi surface, yielding a quasiclassical Green's function
$\check g(i\omega_n, \boldsymbol{r},\boldsymbol{k}_F) = \mathcal{P} \hat\tau_z\frac{i}{\pi}\int d\xi_k\check G_{SC}(\omega_n,\boldsymbol r,\boldsymbol k)$ \cite{Eilenberger1968,Larkin1969}, where $\boldsymbol{k}_F$ denotes the direction of momenta on the Fermi surface, $\xi_k$ is the electronic dispersion relation, $\hat\tau_z$ is a Pauli matrix in Nambu space, and $\mathcal{P}$ indicates principal-value integration. The quasiclassical Green's function $\check g$ is subject to a normalization condition $\check g(i\omega_n, \boldsymbol{r},\boldsymbol{k}_F)^2=\check1$.
Disorder averaging for scattering off non-magnetic, magnetic and spin-orbit impurities \cite{Alexander1985,Demler1997} is performed with the help of the self-consistent Born approximation and results in self-energy corrections to $\check g$.

Further simplification is possible in the dirty limit when the mean free path associated with scattering off non-magnetic impurities is much smaller than the superconducting coherence length (but still much larger than the Fermi wavelength).
In this case one can expand $\check g(i\omega_n, \boldsymbol{r},\boldsymbol{k}_F)$ up to a linear order in $\boldsymbol{k}_F$ and arrive to the Usadel equation for the isotropic (independent of $\boldsymbol{k}_F$) part of the quasiclassical Green's function $\check g(i\omega_n, \boldsymbol r)$. Throughout the paper we make use of the Usadel equations in the form utilized in Refs.~\cite{Ivanov2006,Aikebaier2019}:
\begin{equation}
	D\boldsymbol\partial\cdot (\check g\boldsymbol\partial\check g)-[\omega_n\hat\tau_z +i\boldsymbol{V}_Z^{SC}\cdot\boldsymbol{\hat\sigma}\hat\tau_z+\Delta\hat\tau_++\Delta^{\ast}\hat\tau_- +\check\Sigma,\check g]=0,
	\label{eq:Usadel_full}
\end{equation}
where the covariant derivative is $\boldsymbol\partial \check X=\boldsymbol\nabla-i[\boldsymbol A\hat\tau_z,\check X]$, $\boldsymbol{A}$ is vector potential, $D$ is a diffusion constant associated with electronic scattering off non-magnetic impurities, $\boldsymbol{V}_Z^{SC}=(V_Z^{SC},0,0)$ is the Zeeman field which we assume is uniform and directed along the $x$ axis, $\Delta$ is the pairing potential, $\boldsymbol{\hat\sigma}$($\boldsymbol{\hat\tau}$) is a set of Pauli matrices in spin (Nambu) space and $\hat\tau_{\pm}=(\hat\tau_x\pm i\hat\tau_y)/2$.
Equation~\eqref{eq:Usadel_full} is written in the Nambu spinor basis $(\psi_{\uparrow},\psi_{\downarrow},-\psi_{\downarrow}^{\dagger},\psi_{\uparrow}^{\dagger})^T$.

The self-energy $\check\Sigma=\check\Sigma_{so}+\check\Sigma_{sf}$ incorporates elastic spin relaxation mechanisms that we consider throughout this work: spin-orbit scattering $\check\Sigma_{so}=\boldsymbol{\hat\sigma} \check g\boldsymbol{\hat\sigma}/(8\tau_{so})$ off heavy ions which preserves time-reversal symmetry and spin-flip scattering $\check\Sigma_{sf}=\boldsymbol{\hat\sigma}\hat\tau_z\check g\hat\tau_z\boldsymbol{\hat\sigma}/(8\tau_{sf})$ off magnetic impurities which breaks time-reversal symmetry.
For convenience, we introduce energy scales $\Gamma_{so/sf}=3/(2\tau_{so/sf})$ associated with these two types of scattering.

In this study we neglect orbital effects of the magnetic field which allows us to set $\boldsymbol{A}=0$ and $\Delta\in\mathbb{R}$. The Usadel equation \eqref{eq:Usadel_full} becomes
\begin{equation}
	D\boldsymbol\nabla\cdot (\check g\boldsymbol\nabla\check g)-[\omega_n\hat\tau_z +i\boldsymbol{V}_Z^{SC}\cdot\boldsymbol{\hat\sigma}\hat\tau_z+\Delta\hat\tau_x +\check\Sigma,\check g]=0.
	\label{eq:Usadel}
\end{equation}
Equation~\eqref{eq:Usadel} can be solved by the following Green's function parametrization in terms of functions $\theta(\omega_n,\boldsymbol r)$ and $\phi(\omega_n,\boldsymbol r)$\cite{Ivanov2006,Aikebaier2019}:
\begin{align}
	\check g(\omega_n,\boldsymbol r)=\hat\tau_z\cos\theta(\cosh\phi+i\hat\sigma_x\tan\theta\sinh\phi)+\nonumber \\
	+\hat\tau_x\sin\theta(\cosh\phi-i\hat\sigma_x\cot\theta\sinh\phi).
	\label{eq:g_parametrization}
\end{align}
Note that the parametrization \eqref{eq:g_parametrization} automatically satisfies the normalization condition $\check g^2=\check1$.
The matrix Eq.~\eqref{eq:Usadel} hence becomes a set of nonlinear differential equations
\begin{widetext}
	\begin{subequations}
		\begin{align}
			D\nabla^2\theta+2\cosh\phi(\Delta\cos\theta-\omega_n\sin\theta)-2V_Z^{SC}\sinh\phi\cos\theta-\frac{\Gamma_{sf}}{6}(2\cosh^2\phi+1)\sin2\theta=0, \label{eq:Usad_alg1} \\
			-D\nabla^2\phi+2\sinh\phi(\Delta\sin\theta+\omega_n\cos\theta)-2V_Z^{SC}\cosh\phi\sin\theta+\left(\frac{2\Gamma_{so}}{3}+\frac{\Gamma_{sf}}{3}\cos2\theta\right)\cosh\phi\sinh\phi=0. \label{eq:Usad_alg2}
		\end{align}
	\end{subequations}
\end{widetext}

Once the quasiclassical Green's function $\check g$ is computed via Eqs.~\eqref{eq:g_parametrization}-\eqref{eq:Usad_alg2}, one can evaluate various physical properties of the superconductor, such as the pairing potential, free energy and density of states.
Reference~\cite{Aikebaier2019} derives expressions for these physical quantities in terms of functions $\theta(\omega_n,\boldsymbol r)$, $\phi(\omega_n,\boldsymbol r)$, and here we present those expressions for the reader's convenience.

First, the pairing potential can be calculated by means of the ``gap equation''
\begin{align}
	\Delta\log(\frac{T}{T_{c0}}) & =2\pi T \sum_{\omega_n>0}\left( \frac{1}{4}\Tr(\hat\tau_x\check g)-\frac{\Delta}{\omega_n}\right)\nonumber \\
	                             & =2\pi T \sum_{\omega_n>0}\left(\cosh\phi\sin\theta-\frac{\Delta}{\omega_n}\right),
	\label{eq:gap_eq}
\end{align}
with $T$ being temperature and $T_{c0}$ denoting critical temperature of the superconductor when no Zeeman field or spin relaxation processes are present.
Importantly, Eq.~\eqref{eq:gap_eq} has to be paired with Eqs.~\eqref{eq:Usad_alg1}-\eqref{eq:Usad_alg2} in order to achieve self-consistency of the calculations.
Second, the free energy density difference between the superconducting and normal state can be obtained as \cite{Aikebaier2019}
\begin{widetext}
	\begin{align}
		f_{sn}=\pi T\nu_0 \sum_{\omega_n>0}  \left\{4\omega_n-2\cosh\phi(2\omega_n\cos\theta+\Delta\sin\theta)+4V_Z^{SC}\sinh\phi\sin\theta+D[\nabla^2\theta-\nabla^2\phi]+\right.\nonumber \\
		+\left. \frac{1}{2}\left[\Gamma_{so}+\Gamma_{sf}-(\Gamma_{so}+\Gamma_{sf}\cos2\theta)\cosh^2\phi-
			\frac{1}{3}(\Gamma_{so}-\Gamma_{sf}\cos2\theta)\sinh^2\phi\right]  \right\},
		\label{eq:fSN}
	\end{align}
\end{widetext}
where $\nu_0$ denotes the normal density of sates at the Fermi level.
The condition $f_{sn}<0$ is necessary to ensure thermodynamic stability of the superconducting phase.
Third, the total density of states can be evaluated using quasiclassical Green's function:
\begin{align}
	\nu & =\frac{1}{8}\nu_0\Re[\Tr(\hat\tau_z\check g|_{\omega_n\to-iE^+})]\nonumber \\
	  & =\frac{1}{2}\nu_0\Re[\cos\theta\cosh\phi|_{\omega_n\to-iE^+}].
	\label{eq:DoS}
\end{align}

In general, to analyze the heterostructure shown in Fig.~\ref{fig:Setup}(a), the Usadel equation \eqref{eq:Usadel} [or its parametrized form of Eqs.~\eqref{eq:Usad_alg1}-\eqref{eq:Usad_alg2}] has to be supplemented with boundary conditions, and the corresponding boundary problem has to be solved.
We take the vacuum boundary condition $\boldsymbol{\partial}\check g|_{\text{SC-SM}}=0$ at the interface with the SM.
This is justified by the small effective transparency of the interface for quasiparticles traveling from the superconductor into the semiconductor: quasiparticles in the superconductor have a much larger Fermi momentum than in the semiconductor.
Only quasiparticles moving with a small momentum parallel to the interface can tunnel from the superconductor to the semiconductor, but strong disorder in the superconductor randomizes the momentum direction, resulting in a low-probability of tunneling \cite{Kiendl2019}.
Note that electrons from the semiconductor have a high probability of tunneling into the superconductor and reflecting back as a hole, providing Andreev scattering.
Additional corrections in the very strong tunneling regime involving coherent tunneling and disorder scattering in the superconductor can generate additional subgap states at finite magnetic fields \cite{Liu2018}; we do not consider this regime.
In the absence of orbital effects the vacuum boundary condition becomes the free boundary condition $\boldsymbol{\nabla}\check g|_{\text{SC-SM}}=0$.

The boundary between the superconductor and the magnetic insulator has to be supplemented with an appropriate boundary condition as well.
General spin-dependent boundary conditions for the isotropic superconductor Green's function have been derived in Ref.~\cite{Cottet2009}.
In the case of a boundary between magnetic insulator and thin superconductor (with thickness much smaller than the coherence length $d_{SC}\ll\xi$), the authors of Ref.~\onlinecite{Cottet2009} showed that effects of the magnetic insulator on the superconductor can be described by a uniform effective Zeeman field $V^Z_{eff}\propto d_{SC}^{-1}$ and magnetic scattering induced in the superconductor.
Therefore, given uniformity of the effects induced in the SC by the MI and the free boundary with the SM, we neglect spatial dependence of the SC Green's function $\check g(\omega_n,\boldsymbol r)\to\check g(\omega_n)$ [and correspondingly $\theta(\omega_n,\boldsymbol r)$, $\phi(\omega_n,\boldsymbol r)$].
Consequently, the diffusion constant $D$ drops out of the Usadel equation, and Eqs.~\eqref{eq:Usad_alg1}-\eqref{eq:Usad_alg2} simplify into a set of nonlinear algebraic equations~\footnote{SC-MI bilayers with superconductors of thickness comparable or greater than the coherence length have been recently studied in Ref.~\cite{Hijano2021}}.
Equations~\eqref{eq:Usad_alg1}-\eqref{eq:DoS} provide a sufficient apparatus to self-consistently calculate the quasiclassical  Green's function of the parent SC and study the combined effect of Zeeman field, magnetic and spin-orbit scattering on the superconducting properties. Our calculation is schematically represented in Fig.~\ref{fig:Setup}(b). The spin-orbit scattering is introduced phenomenologically to our model. It can either be intrinsic to the superconductor, for example, as it happens in Pb, or result from scattering off heavy ions.

It is worth mentioning here the crucial difference between clean and dirty superconductors.
In the case of a clean SC, Ref.~\cite{Tokuyasu1988} showed that effects of the MI on the SC are distinct from those generated by the external Zeeman field.
In particular, the transition of the clean SC adjacent to the MI into a normal state can be second order, as opposed to the strictly first order transition in the presence of the Zeeman field.
However, an experiment from Ref.~\cite{Hao1991} demonstrated inadequacy of assuming clean Al when describing EuS-Al bilayers and indicated that dirty Al should be considered instead.
Later on, Ref.~\cite{Cottet2009} showed microscopically that, in the dirty limit, the impact of the MI on the SC is in fact equivalent to that of a Zeeman field and magnetic scattering.

\subsection{\label{sec:sc-sm_proximity}Superconductor-semiconductor proximity effect}

Once the quasiclassical Green's function of the parent SC is calculated, one can analyze the SC-SM proximity effect and investigate emergence of the topological phase in the heterostructure.
The proximity effect arises due to electron tunneling between the superconductor and the semiconductor.
Ignoring irreducible contributions in electron tunneling which can be shown to be much smaller than reducible ones \cite{Stanescu2011, Potter2011, Lutchyn2012}, the proximity effect can be described by the disorder-averaged SC Green's function, and superconducting degrees of freedom in the system can be integrated out.
As a result, effects of the parent SC on the SM can be fully incorporated into the interface self-energy $\check\Sigma(\omega)$\cite{Stanescu2011}.
Assuming spin-independent SC-SM electron tunneling \footnote{Spin-dependent tunneling between the SC and SM has been recently considered in Ref.~\cite{Maiani2020}}, the interface self-energy reads
\begin{equation}
	\check\Sigma(\omega)=|t|^2\nu_0\int d\xi_k \check G_{SC}(\xi_k,\omega)
	\label{eq:self-energy}
\end{equation}
with $|t|$ being the tunneling amplitude.
Note that $\check\Sigma(\omega)$ in Eq.~\eqref{eq:self-energy} does not depend on the Fermi momentum direction.
Recalling the definition of the isotropic quasiclassical Green's function, one can write
\begin{equation}
	\check\Sigma(\omega)=-i\gamma\hat\tau_z\check g(\omega_n)|_{\omega_n\to -i\omega},
	\label{eq:self-energy1}
\end{equation}
where the SC-SM coupling $\gamma=\pi |t|^2\nu_0$ has been introduced.

The Green's function of the quasi-1D SM nanowire can be written as \cite{Stanescu2011}
\begin{align}
	\check G^{-1}(k,\omega )=\omega - V_Z^{SM}\hat\sigma_x-[\xi_{k}+\alpha_R k\hat\sigma_y]\hat\tau_z - \check\Sigma(\omega).
	\label{eq:GF_NW1}
\end{align}
Here $V_Z^{SM}$ is the Zeeman field induced along the direction of the nanowire, for example due to magnetic proximity from the adjacent magnetic insulator and/or external magnetic field, $\xi_{k}=k^2/2m^{\ast}-\mu$ with $m^{\ast}$ and $\mu$ being the SM effective electron mass and chemical potential, respectively, $\alpha_R$ is Rashba spin-orbit coupling, and $\check\Sigma(\omega)$ is given by Eq.~\eqref{eq:self-energy1}.
Writing $\check g(\omega_n)$ as in Eq.~\eqref{eq:g_parametrization} gives
\begin{align}
	&\check G^{-1}(k,\omega)=\omega +i\gamma\cosh\phi\cos\theta- \nonumber \\
	 &-(V_Z^{SM}+\gamma\sinh\phi\sin\theta)\hat\sigma_x-(\xi_k+\alpha_R k\hat\sigma_y)\hat\tau_z- \nonumber \\
	 &-\gamma\cosh\phi\sin\theta\hat\tau_y+i\gamma\sinh\phi\cos\theta\hat\tau_y\hat\sigma_x,
	\label{eq:GF_NW_final}
\end{align}
where $\phi(\omega)$, $\theta(\omega)$ are analytically continued into the real time domain via $\omega_n\to-i\omega$.
Equation~\eqref{eq:GF_NW_final} shows that proximity to the SC induces four extra terms in the nanowire Green's function: frequency shift $\propto \cosh\phi\cos\theta$, Zeeman energy $\propto \sinh\phi\cos\theta$, spin-singlet even-frequency pairing $\propto \cosh\phi\sin\theta$ and spin-triplet odd-frequency pairing $\propto \sinh\phi\cos\theta$.
In the absence of spin-orbit and magnetic scattering in the SC, these proximity-induced terms can be calculated analytically.
Appendix~\ref{sec:Zeeman_analyt} presents the corresponding expressions.

The low-energy spectrum of the system can be determined by computing poles of the Green's function \eqref{eq:GF_NW_final} from
\begin{equation}
	\det[\check G^{-1}(k,\omega)]=0.
	\label{eq:det}
\end{equation}
To this end, one first calculates the self-consistent pair potential and determines the Clogston limit using Eqs.~\eqref{eq:Usad_alg1}-\eqref{eq:fSN}.
Next, the Usadel equations \eqref{eq:Usad_alg1}-\eqref{eq:Usad_alg2} are solved once again, this time in the real time domain using the self-consistent value of the pair potential.
Then, the obtained real time quasiclassical Green's function parametrized through $\phi(\omega)$, $\theta(\omega)$ is plugged into Eq.~\eqref{eq:GF_NW_final}, which produces expressions for the proximity-induced terms in the Green's function.
Finally, the low energy spectrum of the system is inferred by solving Eq.~\eqref{eq:det}.

The spectrum can be used to directly identify topological phase transitions (critical points are signified by gap closure and reopening) and the topological gap.

\section{Results\label{sec:results}}

In the following we first review the results of calculations for the parent SC in isolation. In particular, we analyze the dependence of the superconducting density of states, pair potential and critical temperature on Zeeman field and spin-orbit and magnetic scattering.
Then we consider the SC-SM proximity effect and calculate the dependence of the topological phase transition on these parameters.
Finally, the dependence of the topological gap on external magnetic field for various values of Zeeman field and/or spin-orbit and magnetic scattering in the SC is demonstrated.

\subsection{\label{sec:SC_calc}Properties of the parent superconductor}

\begin{figure*}[ht!]
	\includegraphics[width = 0.68\columnwidth]{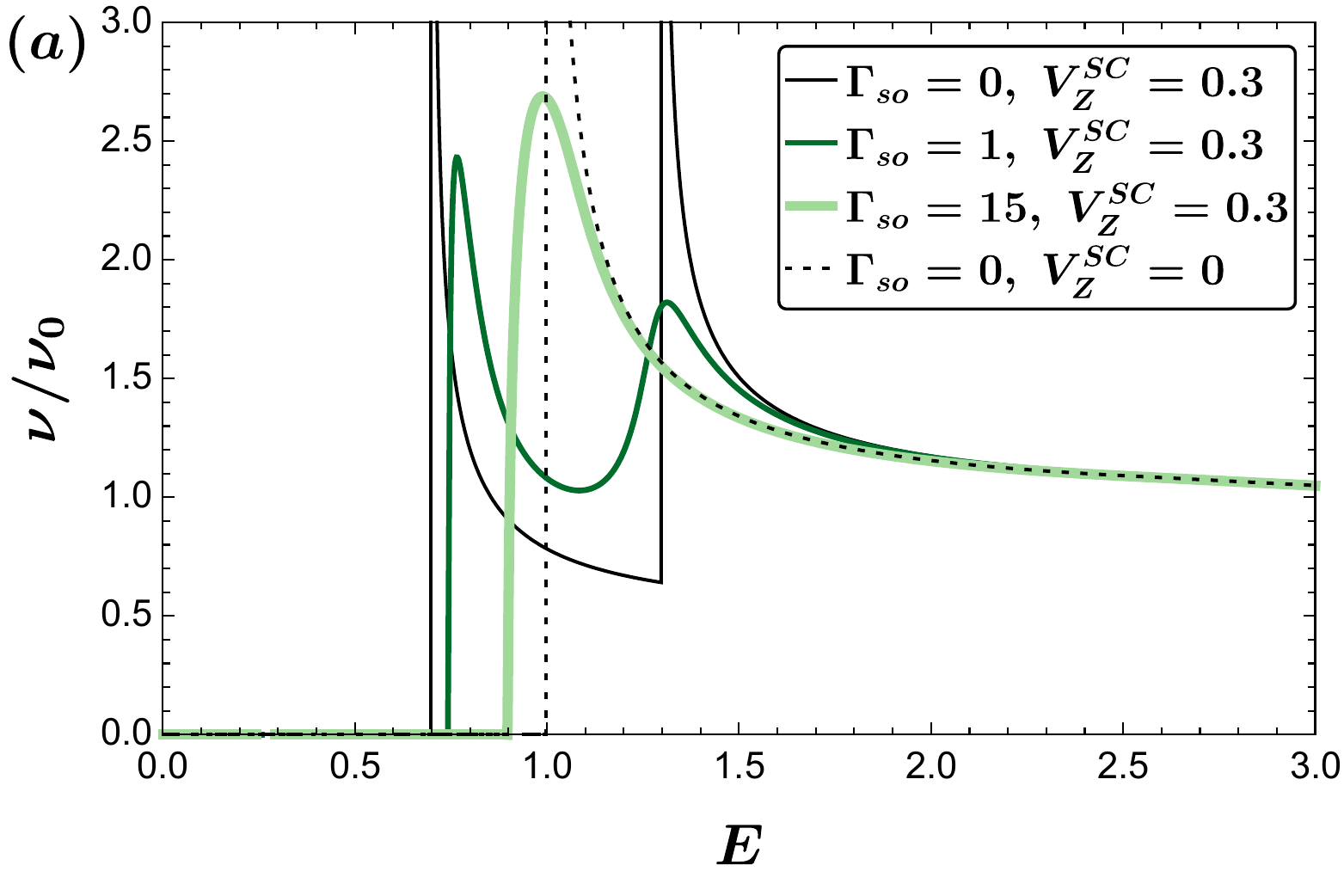}
	\includegraphics[width = 0.68\columnwidth]{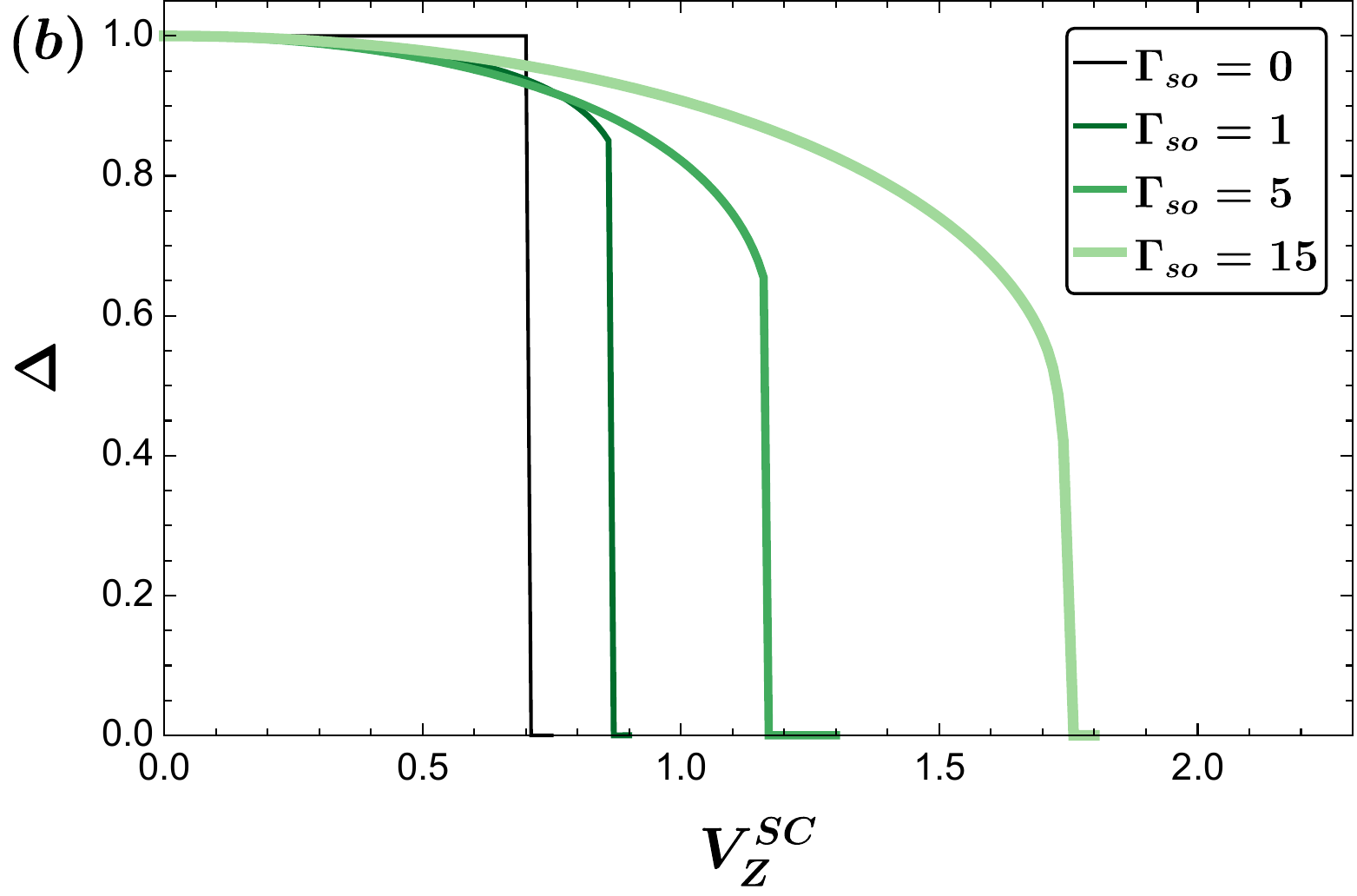}
	\includegraphics[width = 0.68\columnwidth]{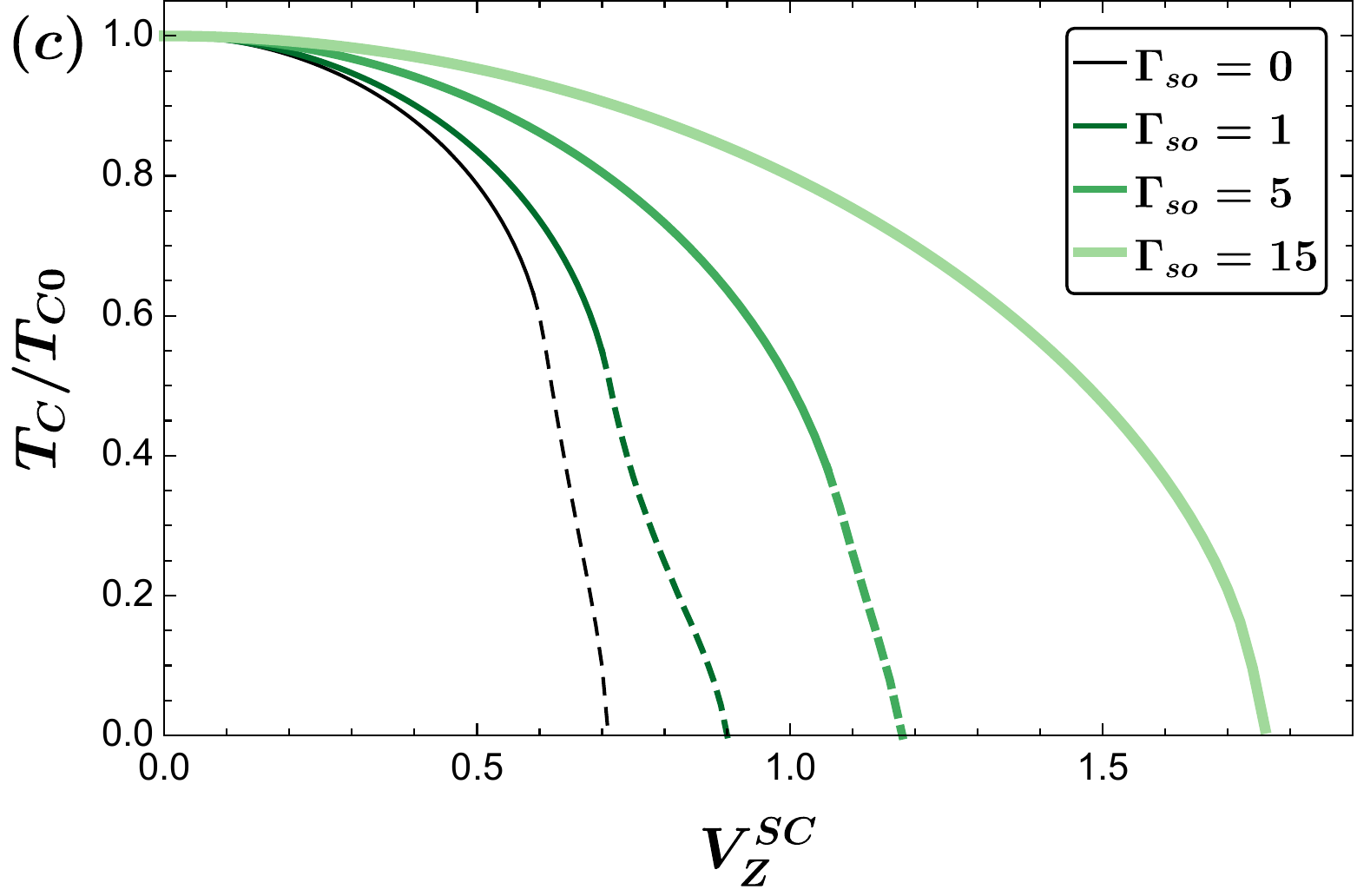}
	\includegraphics[width = 0.68\columnwidth]{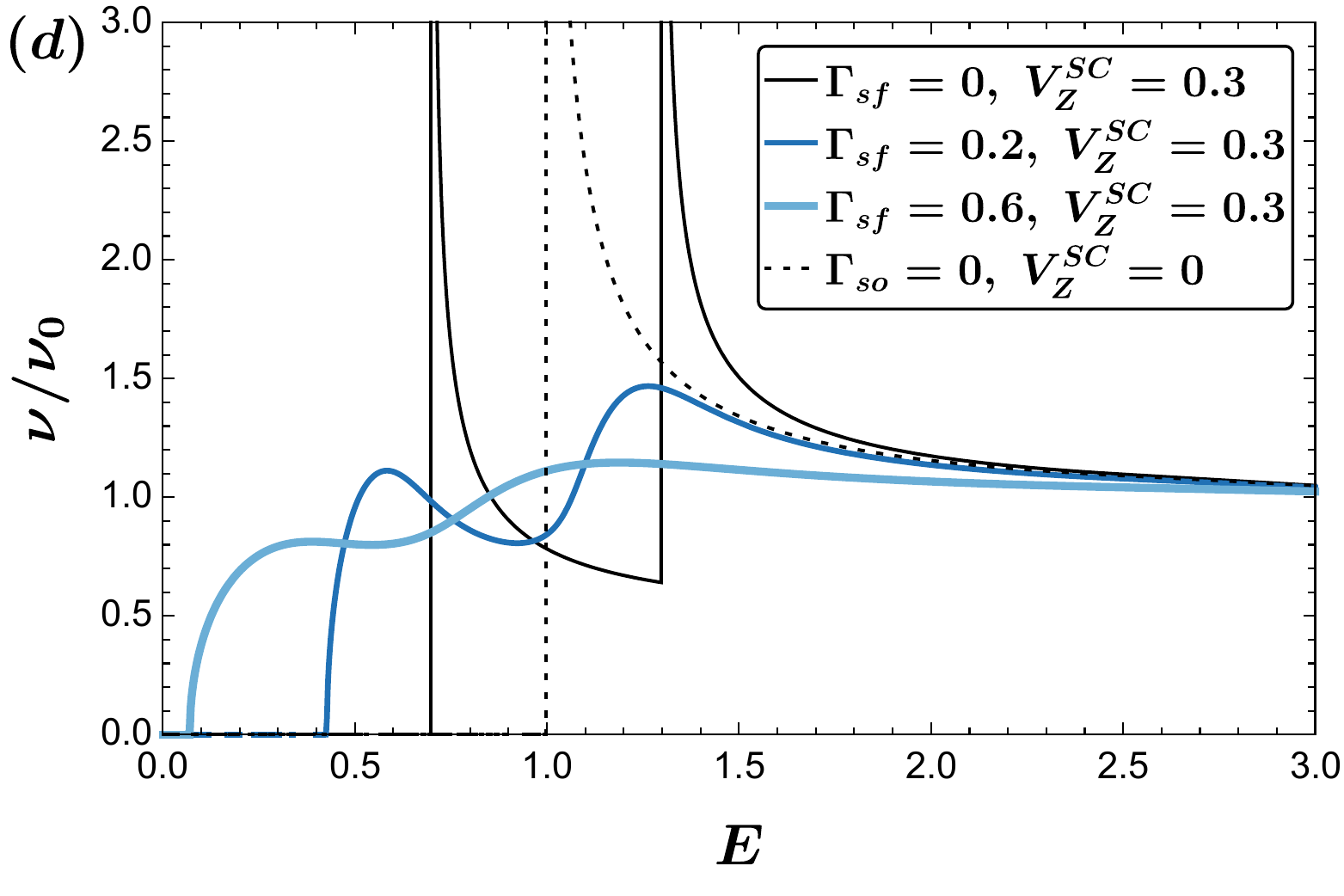}
	\includegraphics[width = 0.68\columnwidth]{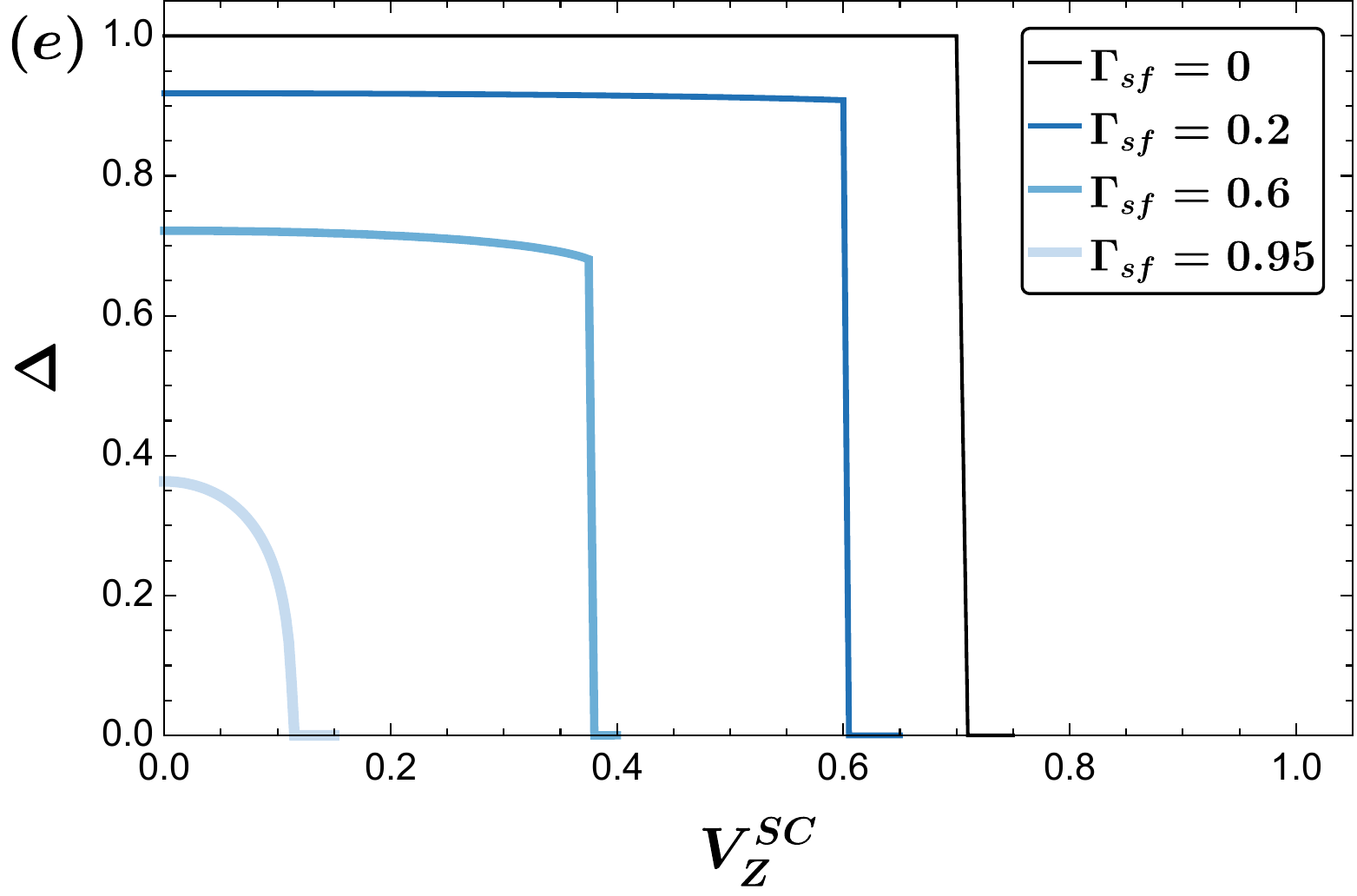}
	\includegraphics[width = 0.68\columnwidth]{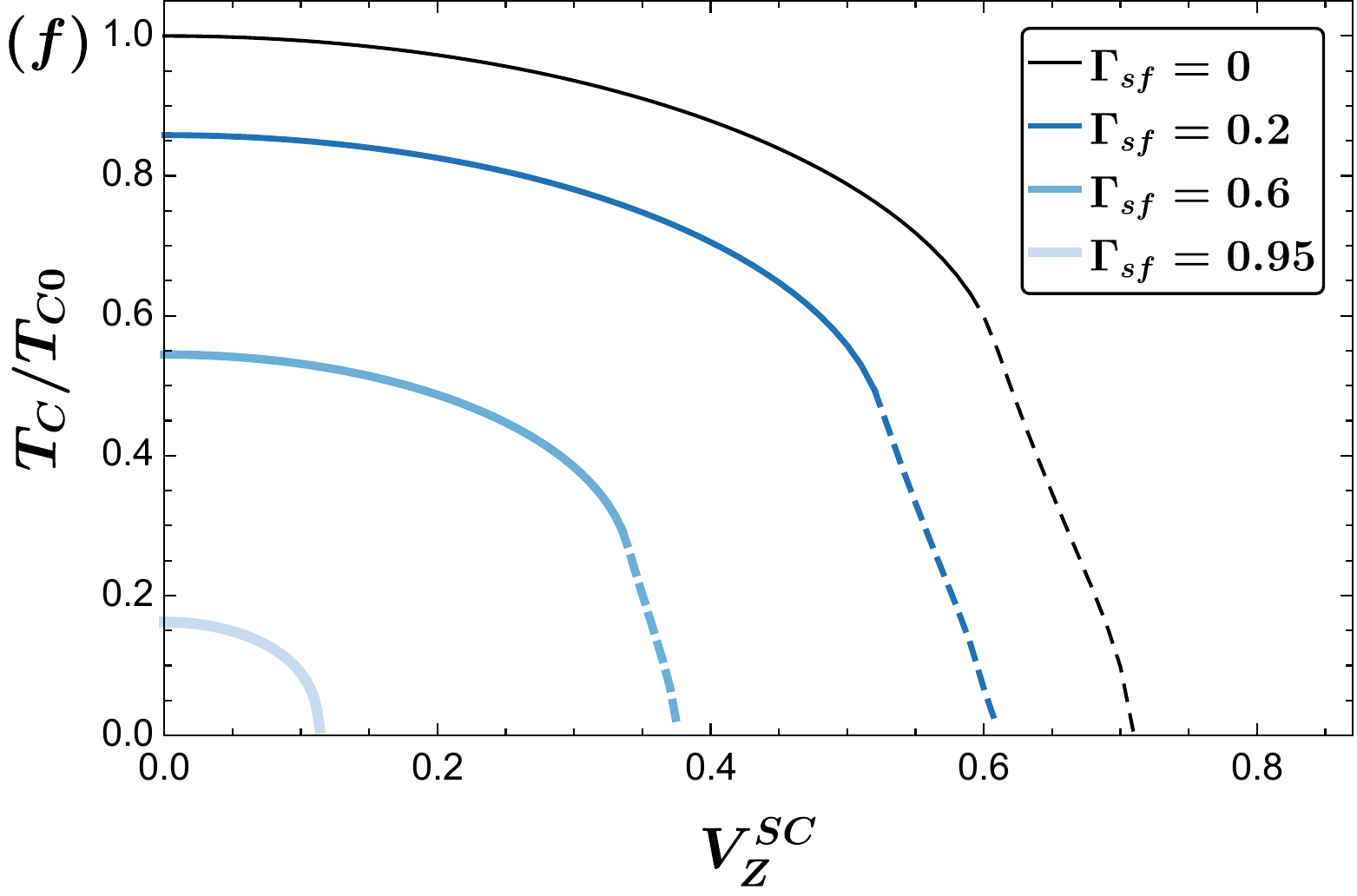}
	\vspace{-1em}
	\caption{Dependence of density of states (left column), pairing potential (middle column) and critical temperature (right column) of the SC on Zeeman field $V_Z^{SC}$ for various values of spin-orbit (upper row, $\Gamma_{so}$) and magnetic (lower row, $\Gamma_{sf}$) scattering energy. Magnetic scattering in (a)-(c) and spin-orbit scattering in (d)-(f) is set to zero. For (a)-(b) and (d)-(e), temperature is fixed to $T/T_{c0}=0.01$. Solid (dashed) lines in (c) and (f) indicate second (first) order transition. All energies in (a)-(f) are measured in units of the bare superconducting gap $\Delta_{00}$.}
	\label{fig:SC_so_sf}
\end{figure*}

We begin with an analysis of how Zeeman energy and spin-orbit scattering affect the SC. Although the results in this section are established~\cite{Sarma1963,Bruno1973,Bergeret2018,Heikkila2019}, we reproduce them here both as a validation of our methodology and to make the presentation self-contained.
Figure~\ref{fig:SC_so_sf}(a) depicts the SC density of states of Eq.~\eqref{eq:DoS} plotted for a fixed value of Zeeman field and various values of spin-orbit scattering energies.
Throughout energy is measured in units of the bare gap $\Delta_{00}$ of the SC when no Zeeman field or spin relaxation processes are present.
Figure~\ref{fig:SC_so_sf}(a) demonstrates that the Zeeman energy splits the density of states in the superconductor into two spin bands.
Spin-orbit scattering reduces the spin splitting in the SC formed by the Zeeman field and eventually merges the two spin-resolved peaks in the density of states into one; in the limit of infinite spin-orbit scattering a single-peak BCS density of states is recovered \cite{Bruno1973}.
The pairing potential is likewise affected by spin-orbit scattering, as illustrated in Fig.~\ref{fig:SC_so_sf}(b) where we plot it as a function of Zeeman field for several values of $\Gamma_{so}$.
In the absence of spin-orbit scattering, the pairing potential is constant as a function of $V_Z^{SC}$ up to a value of $V_Z^{SC}=1/\sqrt{2}$ where it vanishes and the system undergoes a first order transition into the normal state.
This critical value of Zeeman field is the Clogston limit~\cite{Chandrasekhar1962,Clogston1962,Sarma1963} representing the maximum paramagnetic spin splitting that the SC can withstand.
Adding spin-orbit scattering to the system pushes the Clogston limit to higher critical values of $V_Z^{SC}$; see Fig.~\ref{fig:SC_so_sf}(b).
The pair potential in turn decreases as a function of $V_Z^{SC}$ when $\Gamma_{so}\neq 0$ and for values of $\Gamma_{so}\gtrsim 15$ the transition to the normal state becomes second order.
Lastly, we consider how the critical temperature $T_c$ is impacted by both Zeeman field and spin-orbit scattering in Fig.~\ref{fig:SC_so_sf}(c).
At finite Zeeman splitting spin-orbit scattering increases the critical temperature; just as in Fig.~\ref{fig:SC_so_sf}(b), spin-orbit scattering increases the Clogston limit, while values of spin-orbit scattering larger than a certain threshold result in the transition switching from first to second order.

Now we consider the combined effect of Zeeman energy and magnetic scattering on the SC in Fig.~\ref{fig:SC_so_sf}(d)-\ref{fig:SC_so_sf}(f).
We present the density of states, plotted for a fixed value of Zeeman field and several values of magnetic scattering energies, in Fig.~\ref{fig:SC_so_sf}(d).
Magnetic scattering smears out the density of states peaks and reduces the excitation gap.
For sufficiently large values of $\Gamma_{sf}$, the SC becomes gapless~\cite{Tinkham2004}.
Moving on to the analysis of the pairing potential, Fig.~\ref{fig:SC_so_sf}(e) depicts its dependence on Zeeman field and magnetic scattering.
We observe that the presence of magnetic scattering in the SC reduces the pairing potential and, unlike spin-orbit scattering, decreases the Clogston limit.
At the same time, similar to spin-orbit scattering, adding a sufficiently large $\Gamma_{sf}$ switches the order of the transition from first to second.
Figure~\ref{fig:SC_so_sf}(f), which presents the dependence of the critical temperature on Zeeman field and magnetic scattering, supports the above conclusions.
The opposite behavior compared to spin-orbit scattering is observed, i.e., increasing the magnitude of magnetic scattering that breaks time-reversal symmetry reduces the critical temperature at all values of induced Zeeman splitting.

\subsection{\label{sec:top_phase_transition}Topological phase transition}

Having calculated self-consistent values of the pair potential in the parent SC, we now analyze the emergence of the topological phase in the proximitized nanowire.
Importantly, the frequency dependence of the Green's function in Eq.~\eqref{eq:GF_NW_final} is irrelevant to the topological phase transition: at the critical point the energy gap at $k=0$ closes, which enables one to identify the transition by setting $\omega_n=0$ in Eqs.~\eqref{eq:Usad_alg1}-\eqref{eq:Usad_alg2} and $k=\omega=0$ in Eq.~\eqref{eq:GF_NW_final} as long as the correct self-consistent value of the pair potential $\Delta=\Delta(V_Z^{SC},\Gamma_{so},\Gamma_{sf})$ is given.
In this case, Eq.~\eqref{eq:Usad_alg1} gives $\theta=\pi/2$ while Eq.~\eqref{eq:Usad_alg2} reads
\begin{align}
	\Delta\sinh\phi &- V_Z^{SC}\cosh\phi+ \nonumber\\
	&+\frac{1}{3}\left(\Gamma_{so}-\frac{\Gamma_{sf}}{2}\right)\cosh\phi\sinh\phi=0.
	\label{eq:Usad_transition}
\end{align}
At the same time, the Green's function \eqref{eq:GF_NW_final} becomes
\begin{align}
	\check G^{-1}(k=0,\omega=0)= -(V_Z^{SM}+\gamma\sinh\phi)\hat\sigma_x -\gamma\cosh\phi\hat\tau_y,
	\label{eq:GF_NW_k0w0}
\end{align}
where for simplicity we set chemical potential in the nanowire to zero, $\mu=0$.
Note that the induced Zeeman energy (spin-singlet pairing) in the nanowire is equal to $\gamma\sinh\phi$ ($\gamma\cosh\phi$) while odd-frequency spin-triplet pairing vanishes.
From Eq.~\eqref{eq:GF_NW_k0w0} we identify the minimum Zeeman field in the SM necessary to create the topological phase, $V_{Z,c}^{SM}$, as
\begin{equation}
	V_{Z,c}^{SM}= \gamma(\cosh\phi-\sinh\phi).
	\label{eq:VZSMc_def}
\end{equation}
Rewriting Eq.~\eqref{eq:Usad_transition} in terms of this quantity gives
\begin{align}
	\left(\Delta-V_Z^{SC}\right)\left(V_{Z,c}^{SM}/\gamma\right)-\left(\Delta+V_Z^{SC}\right)\left(V_{Z,c}^{SM}/\gamma\right)^3+\nonumber\\
	+\frac{\Gamma}{6}\left[1-\left(V_{Z,c}^{SM}/\gamma\right)^4\right]=0,
	\label{eq:VZSMc_eq}
\end{align}
where we denoted $\Gamma\equiv\Gamma_{so}-\Gamma_{sf}/2$.
Note that $\Gamma$ in general does not fully incorporate effects of spin relaxation processes on the topological critical point because the value of the self-consistent pair potential $\Delta=\Delta(V_Z^{SC},\Gamma_{so},\Gamma_{sf})$ depends on these processes as well.
Although an analytic solution to the quartic equation \eqref{eq:VZSMc_eq} exists, in general it is cumbersome and we do not present it here.
Instead, we build intuition by considering limiting behavior of Eq.~\eqref{eq:VZSMc_eq}.
First, in the limit of $\Gamma=0$, which corresponds either to the absence of spin relaxation processes $\Gamma_{so}=\Gamma_{sf}=0$ or to the case when $\Gamma_{so}=\Gamma_{sf}/2$, we find that a minimum Zeeman field of
\begin{equation}
	\frac{V_{Z,c}^{SM}}{\gamma}=\sqrt{\frac{\Delta-V_Z^{SC}}{\Delta+V_Z^{SC}}}
	\label{eq:VZSMc_Gso0}
\end{equation}
is required in the SM in order to induce topological superconductivity.
For small but non-zero values of $\Gamma$, $|\Gamma|\ll \left(\Delta^2-\left(V_Z^{SC}\right)^2\right)^{3/2}/\left(V_Z^{SC}\Delta\right)$, perturbative corrections to $V_{Z,c}^{SM}$ of Eq.~\eqref{eq:VZSMc_Gso0} can be calculated. Up to first order in $\Gamma$ we obtain
\begin{align}
	\frac{V_{Z,c}^{SM}}{\gamma}=\sqrt{\frac{\Delta-V_Z^{SC}}{\Delta+V_Z^{SC}}}+\frac{\Delta V_Z^{SC}}{3(\Delta-V_Z^{SC})(\Delta+V_Z^{SC})^2}\Gamma+\nonumber\\
	+O(\Gamma^2).
	\label{eq:VZSMc_Gso_small}
\end{align}
In the absence of spin-orbit and magnetic scattering, Eq.~\eqref{eq:VZSMc_Gso0} demonstrates that it is impossible to close the gap without an additional Zeeman field in the SM: in this case the topological phase requires $V_Z^{SC} > \Delta$ which is prohibited by the Clogston limit.
Adding a small spin-orbit scattering, which leads to a small positive $\Gamma$, does not improve the situation.
On the contrary, the corresponding correction in Eq.~\eqref{eq:VZSMc_Gso_small} increases value of the critical Zeeman field.
This behavior is a manifestation of the fact that spin-orbit scattering quenches spin splitting in the SC; see Fig.~\ref{fig:SC_so_sf}(a) and the corresponding discussion in Section~\ref{sec:SC_calc}.
Due to this fact, the effective Zeeman energy transferred from the SC to the SM is decreased by the presence of spin-orbit scattering in the superconductor.
On the other hand, adding purely magnetic scattering, which leads to negative $\Gamma$, reduces the critical SM Zeeman field for a fixed value of $V_Z^{SC}$.
However, magnetic scattering also suppresses the Clogston limit as has been discussed in Section~\ref{sec:SC_calc}, so that the maximum $V_Z^{SC}$ that the parent SC can sustain is smaller.
For this reason, adding magnetic scattering does not assist in reducing $V_{Z,c}^{SM}$.
We show this below when we solve Eq.~\eqref{eq:VZSMc_eq} numerically for self-consistent $\Delta$ and general values of $V_Z^{SC},\Gamma_{so},\Gamma_{sf}$.

Next, we consider the opposite limit of infinite spin-orbit scattering, $\Gamma_{so}\approx\Gamma\to\infty$, or no Zeeman splitting in the SC, $V_Z^{SC}=0$.
In both of these limits, Eq.~\eqref{eq:VZSMc_eq} yields $V_{Z,c}^{SM}=\gamma$ regardless of the values of other parameters.
Note that $V_{Z,c}^{SM}=\gamma$ is a familiar result for the topological criterion at zero chemical potential \cite{Stanescu2011}.
Expanding the solution of Eq.~\eqref{eq:VZSMc_eq} near $V_{Z,c}^{SM}=\gamma$, we can find corrections for finite but large $\Gamma_{so}$ or nonzero $V_Z^{SC}$:
\begin{equation}
	\frac{V_{Z,c}^{SM}}{\gamma}= 1-\frac{3V_Z^{SC}}{3\Delta+6V_Z^{SC}+\Gamma_{so}}+\cdots,
	\label{eq:VZSMc_Gso_VZSC}
\end{equation}
which is valid as long as $3V_Z^{SC}\ll 3\Delta+6V_Z^{SC}+\Gamma_{so}$.
Further expanding Eq.~\eqref{eq:VZSMc_Gso_VZSC} in the limit of large spin-orbit scattering $\Gamma_{so}\gg V_Z^{SC},\Delta$ leads to
\begin{equation}
	\frac{V_{Z,c}^{SM}}{\gamma}= 1-\frac{3V_Z^{SC}}{\Gamma_{so}}+O\left(\frac{1}{\Gamma_{so}^2} \right).
	\label{eq:VZSMc_Gso_large}
\end{equation}
Equation~\eqref{eq:VZSMc_Gso_large} shows once again that spin-orbit scattering suppresses Zeeman splitting in the SC.
On the other hand, expanding Eq.~\eqref{eq:VZSMc_Gso_VZSC} in the limit of small SC Zeeman energy $V_Z^{SC}\ll\Gamma_{so},\Delta$ gives
\begin{equation}
	\frac{V_{Z,c}^{SM}}{\gamma}= 1-\frac{3V_Z^{SC}}{3\Delta+\Gamma_{so}}+O\left(\left(V_Z^{SC}\right)^2\right).
	\label{eq:VZSMc_VZSC_small}
\end{equation}

\begin{figure}[t!]
	\includegraphics[width = 0.95\columnwidth]{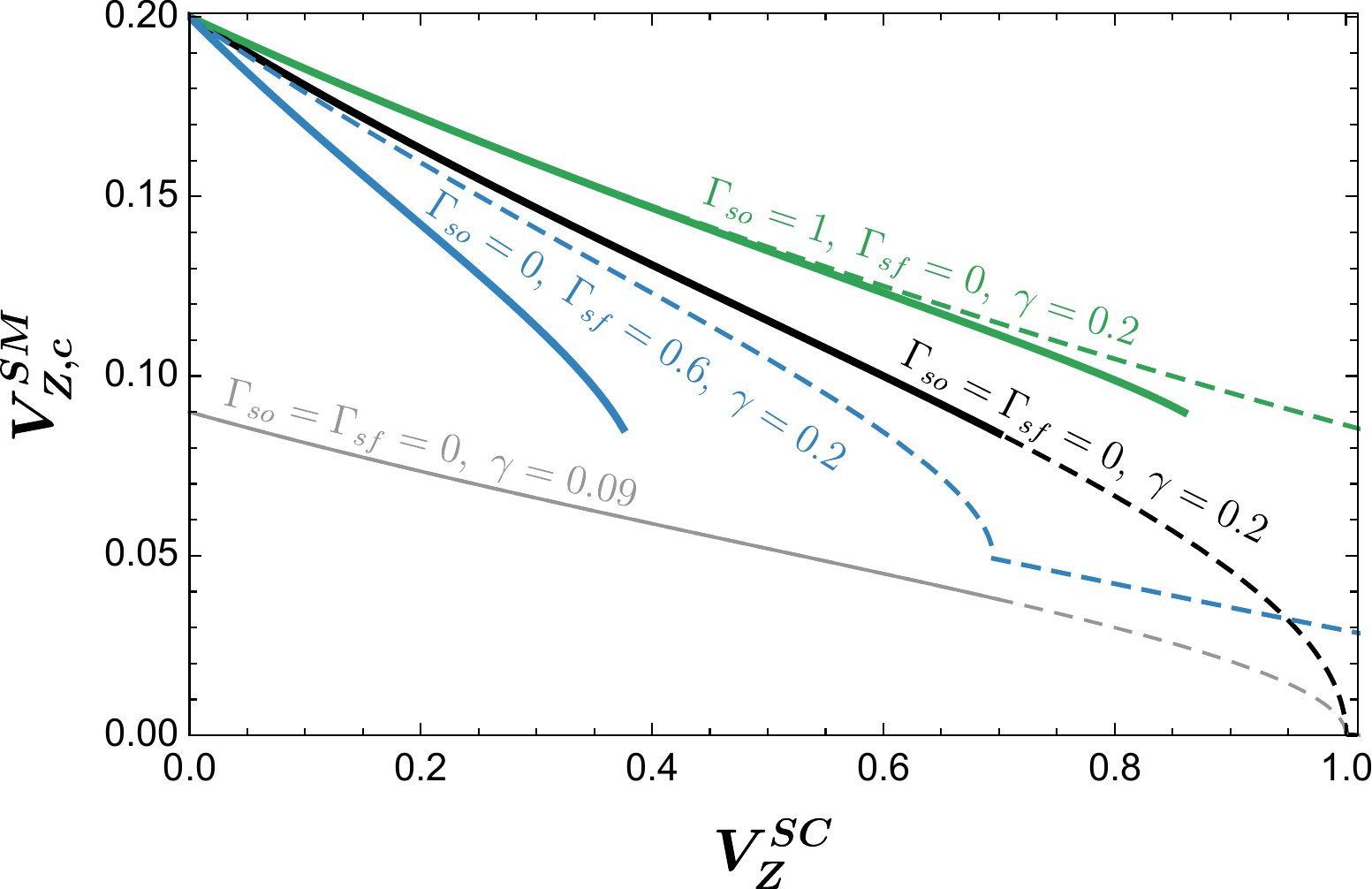}
	\caption{Critical Zeeman energy $V_{Z,c}^{SM}$ required to be added to the nanowire in order to induce topological phase versus Zeeman energy in the parent superconductor $V_Z^{SC}$ for various values of spin-orbit and magnetic scattering and superconductor-semiconductor coupling $\gamma$. Solid lines are calculated self-consistently and terminate at the Clogston limit, while dashed lines are calculated non-self-consistently. Chemical potential in the nanowire is set to zero. All energies are measured in units of the bare superconducting gap $\Delta_{00}$.}
	\label{fig:VZSM_VZSC}
\end{figure}

Figure~\ref{fig:VZSM_VZSC} presents the solution of Eq.~\eqref{eq:VZSMc_eq} for general values of $V_Z^{SC}$ and $\Gamma_{so}$, $\Gamma_{sf}$.
Self-consistent values of the pair potential $\Delta(V_Z^{SC},\Gamma_{so},\Gamma_{sf})$ cannot be in general obtained analytically, so in Fig.~\ref{fig:VZSM_VZSC} we depict the curves $V_{Z,c}^{SM}(V_Z^{SC})$ for two different settings: (1) solid curves represent the situation when numerically calculated values of the self-consistent pair potential $\Delta=\Delta(V_Z^{SC},\Gamma_{so},\Gamma_{sf})$ [Fig.~\ref{fig:SC_so_sf}(b),(e)] are used in Eq.~\eqref{eq:VZSMc_eq}, while (2) dashed curves illustrate the case when Eq.~\eqref{eq:VZSMc_eq} is solved with the non-self-consistent pair potential $\Delta=\Delta_{00}$.
Each solid curve in Fig.~\ref{fig:VZSM_VZSC} terminates at its respective Clogston limit when the parent superconductor transitions into the normal state.
Even though spin-orbit scattering, if present in the SC, can push Clogston limit further --- see Fig.~\ref{fig:SC_so_sf}(b)-\ref{fig:SC_so_sf}(c) and discussion in Section~\ref{sec:SC_calc} --- it also quenches spin splitting in the SC and, correspondingly, Zeeman energy transferred to the SM.
At the same time, magnetic scattering suppresses the Clogston limit as can be seen in Figs.~\ref{fig:SC_so_sf}(e)-\ref{fig:SC_so_sf}(f).
On top of that, magnetic scattering quenches the pair potential in the parent SC and thus has a detrimental effect on the topological gap; see Section~\ref{sec:top_gap}.
For this reason, spin-orbit and magnetic scattering do not assist in creating topological superconductivity in the nanowire.

Therefore, we emphasize again that the topological phase \textit{cannot} be achieved in hybrid heterostructures when the Zeeman field is only induced in the parent superconductor, regardless of whether spin-orbit or magnetic scattering is present in the superconductor.
This conclusion has been stated in recent works by other authors as well~\cite{Woods2020,Maiani2020}, although they did not explicitly consider spin relaxation processes in the SC.

Reduction of the critical field $V_{Z,c}^{SM}$ can be achieved by weakening the SC-SM coupling $\gamma$, see Eq.~\eqref{eq:VZSMc_def} and the grey line in Fig.~\ref{fig:VZSM_VZSC}. However, in the weak coupling regime a smaller coupling leads to a smaller topological gap, see Section~\ref{sec:top_gap}.

Dashed curves in Fig.~\ref{fig:VZSM_VZSC} represent the results of non-self-consistent calculations.
For $\Gamma_{sf}=0$ these curves closely follow the solid self-consistent lines.
However, in the presence of magnetic scattering (blue curve in Fig.~\ref{fig:VZSM_VZSC}) the dashed and the solid curves are positioned considerably off from each other because magnetic scattering substantially reduces the pairing potential; see Fig.~\ref{fig:SC_so_sf}(e).
This reduction cannot be captured by the non-self-consistent pair potential.
Another drawback of the non-self-consistent calculation is that it does not enforce the Clogston limit, and therefore the dashed curves in Fig.~\ref{fig:VZSM_VZSC} do not terminate.
This pathology could lead to incorrect conclusions about the topological phase diagram.
For this reason, we emphasize the importance of self-consistency in the SC Green's function (and subsequent topological phase diagram) calculation if Zeeman energy and spin relaxation mechanisms are present in the superconductor.

\subsection{\label{sec:top_gap}Topological gap}

Beyond the critical point, a crucial property of the topological phase is the spectral gap.
A larger gap enhances the protection of the topological phase against quasiparticle poisoning and disorder.
In this subsection, we calculate the gap in the nanowire and analyze its dependence on Zeeman splitting and spin-orbit and magnetic scattering in the parent SC.
In general, because of the nontrivial frequency dependence of the Green's function \eqref{eq:GF_NW_final}, the energy spectrum of the nanowire has to be computed numerically.
Moreover, in case of spin relaxation processes present in the SC, the Usadel equations \eqref{eq:Usad_alg1}-\eqref{eq:Usad_alg2} are solved numerically as well.
For this reason, we perform a numerical analysis of Eqs.~\eqref{eq:Usad_alg1}-\eqref{eq:Usad_alg2}, \eqref{eq:GF_NW_final}.
Throughout the calculations we set the Rashba coupling in the nanowire to $\alpha=0.2\ \text{eV}\cdot\text{\AA}$ and the effective electron mass to $m^{\ast}=0.02m_0$, where $m_0$ is the electron rest mass.
We continue using the bare gap of the parent SC $\Delta_{00}=0.23\ \text{meV}$ as a unit of energy.

As discussed in Section \ref{sec:top_phase_transition}, a certain Zeeman field $V_{Z,c}^{SM}$ has to be introduced directly to the semiconductor to achieve the topological phase.
Here we consider the case when $V_{Z,c}^{SM}$ is created by applying an external Zeeman field $V_Z^{ext}$ to the entire system.
As stated in Section~\ref{sec:methods}, we ignore orbital effects of the magnetic field.
We assume that there is no coupling between the SM and the MI, although this assertion can be easily adjusted in our framework by using a different parametrization of $V_Z^{ext}$.
The total Zeeman energies of the SC and the SM are $V_{Z}^{SC}=g_{SC}V_Z^{ext}+V_{Z,0}^{SC}$ and $V_Z^{SM}=g_{SM}V_Z^{ext}$, respectively, where $V_{Z,0}^{SC}$ is the MI-induced Zeeman splitting in the SC and we take $g_{SC}=2$ and $g_{SM}=-15$.

\begin{figure}[t!]
	\includegraphics[width = \columnwidth]{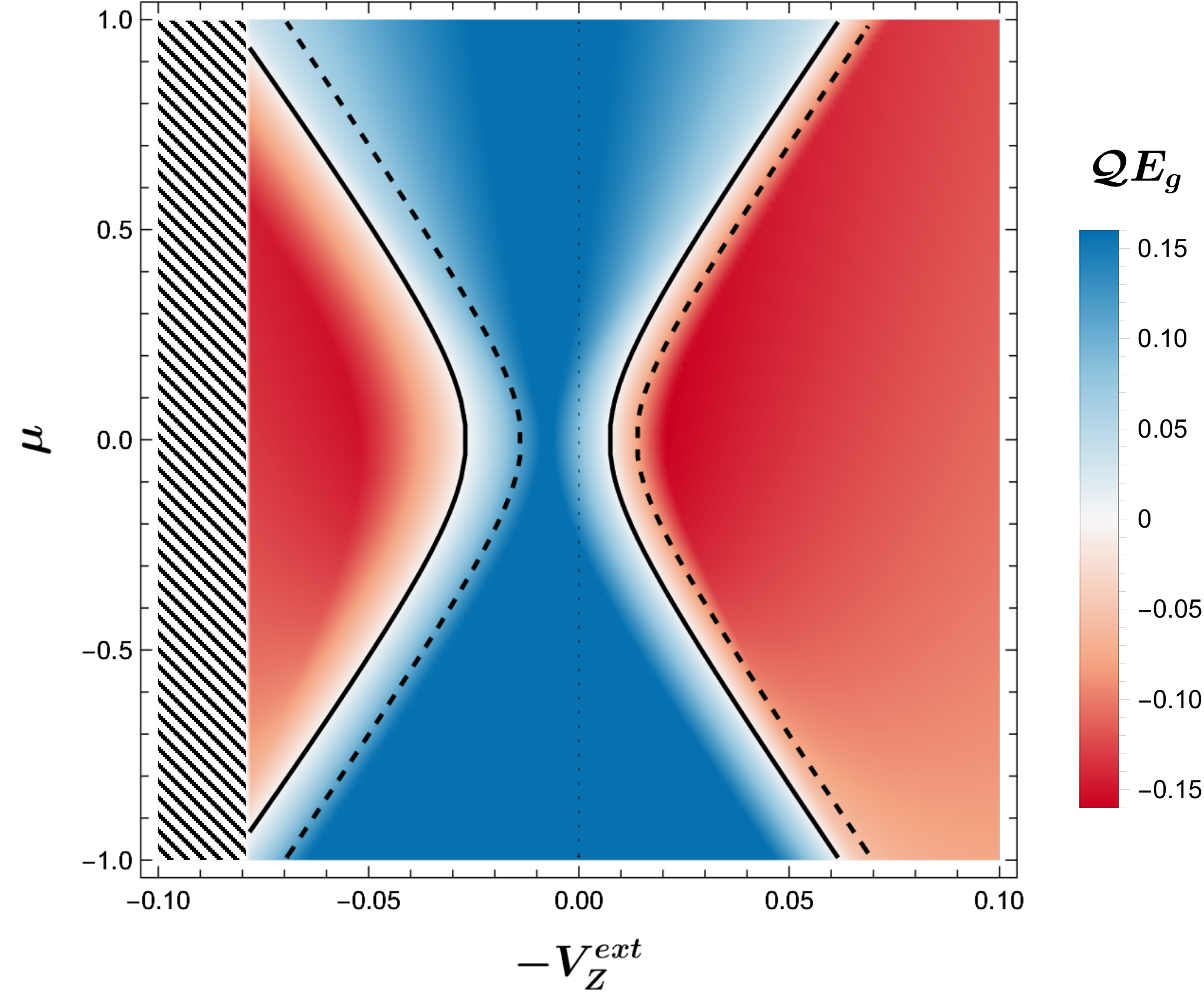}
	\caption{Spectral gap $E_g$ (color) multiplied by the topological invariant $\mathcal{Q}=\pm 1$ as a function of the chemical potential $\mu$ in the semiconductor and the external Zeeman field $V_Z^{ext}$ for $\Gamma_{so}=\Gamma_{sf}=0$, $V_{Z,0}^{SC}=0.55$. The phase boundary is marked by solid black lines. Dashed black lines represent the phase boundary when $V_{Z,0}^{SC}=0$. The hatched region depicts the part of the phase diagram beyond the Clogston limit when superconductivity in the parent SC is destroyed. All energies are measured in units of the parent superconductor's bare gap $\Delta_{00}$.
	}
	\label{fig:phase_diag}
\end{figure}

\begin{figure*}[ht!]
	\includegraphics[width = 0.68\columnwidth]{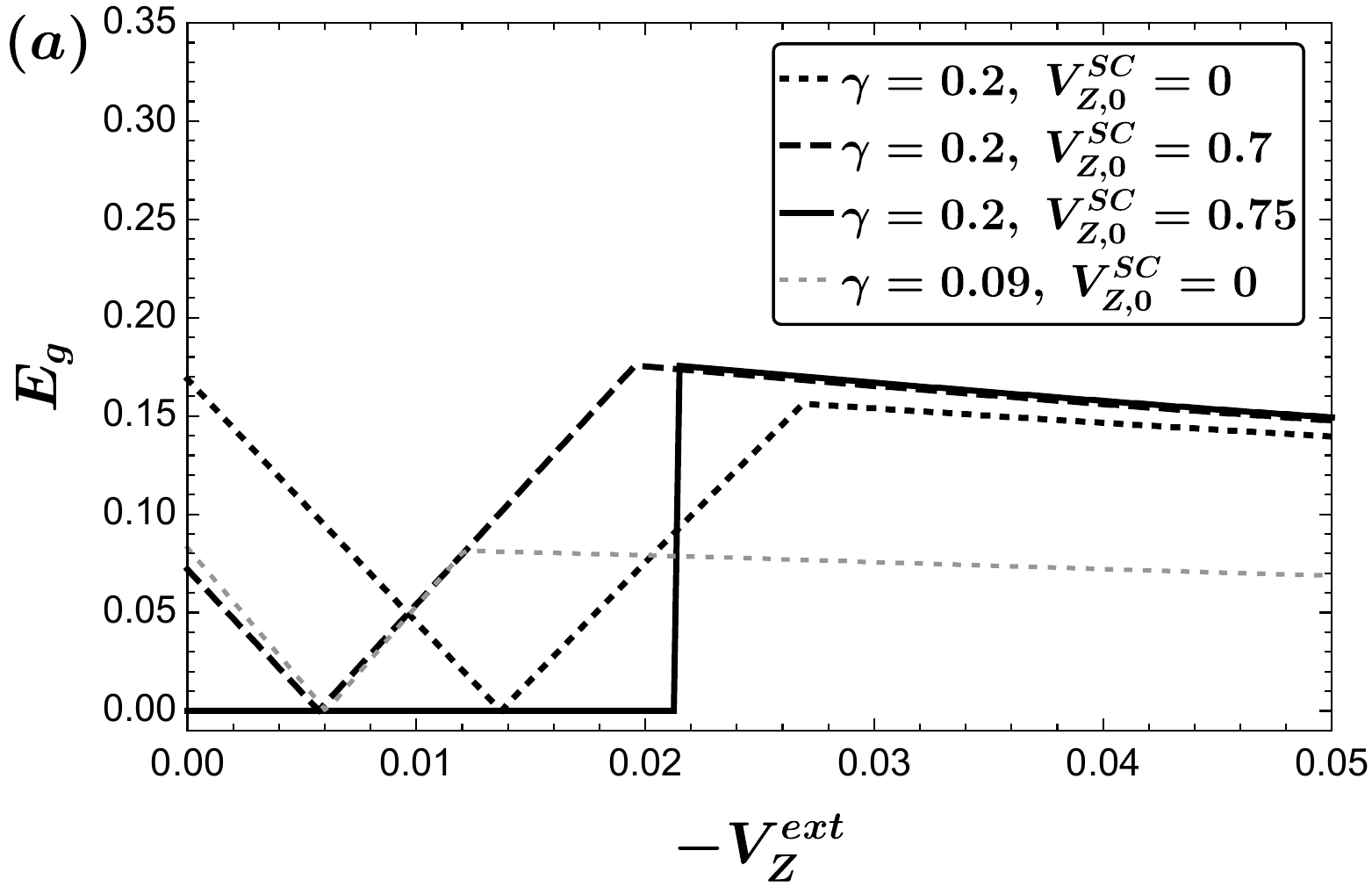}
	\includegraphics[width = 0.68\columnwidth]{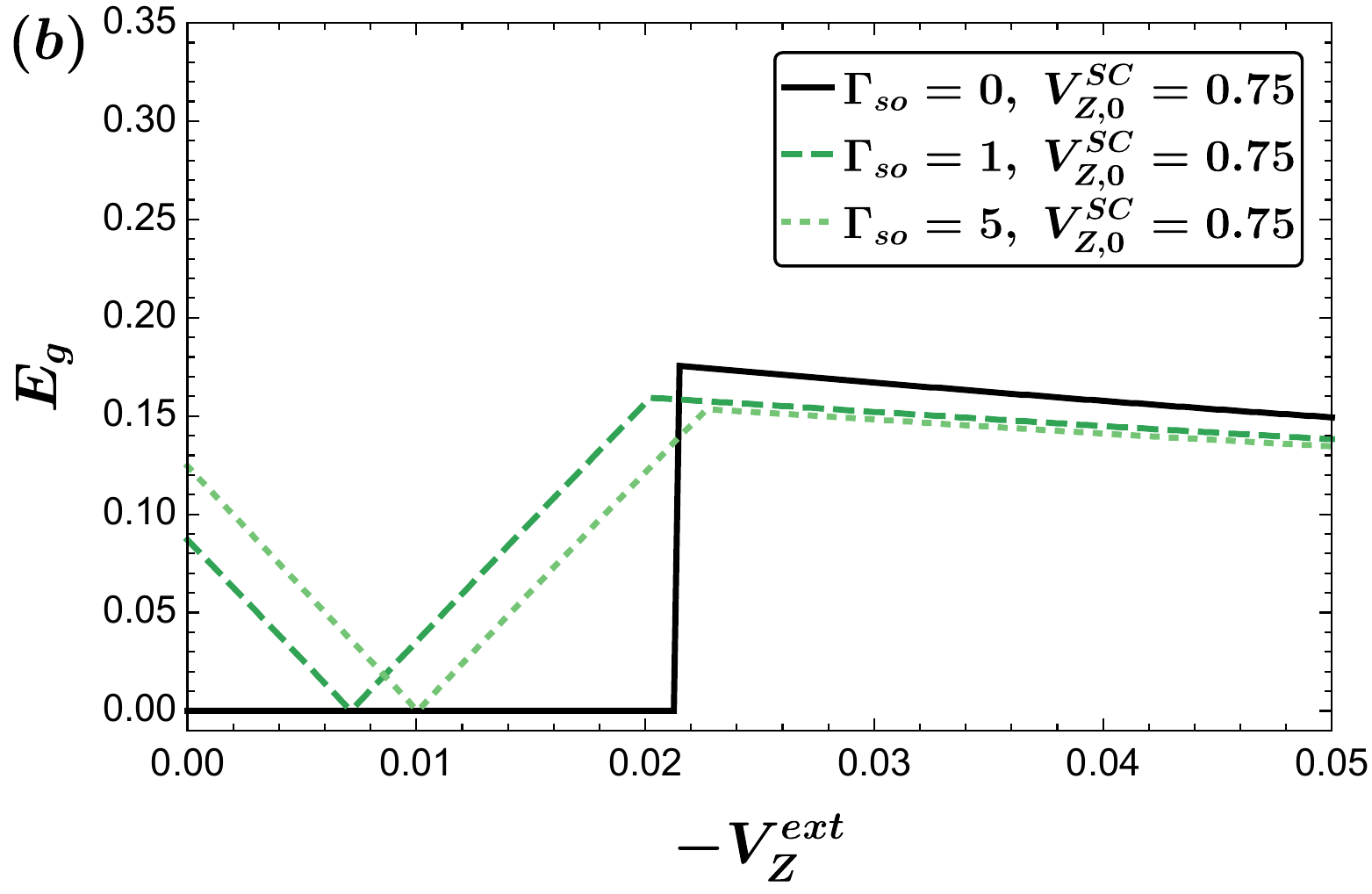}
	\includegraphics[width = 0.68\columnwidth]{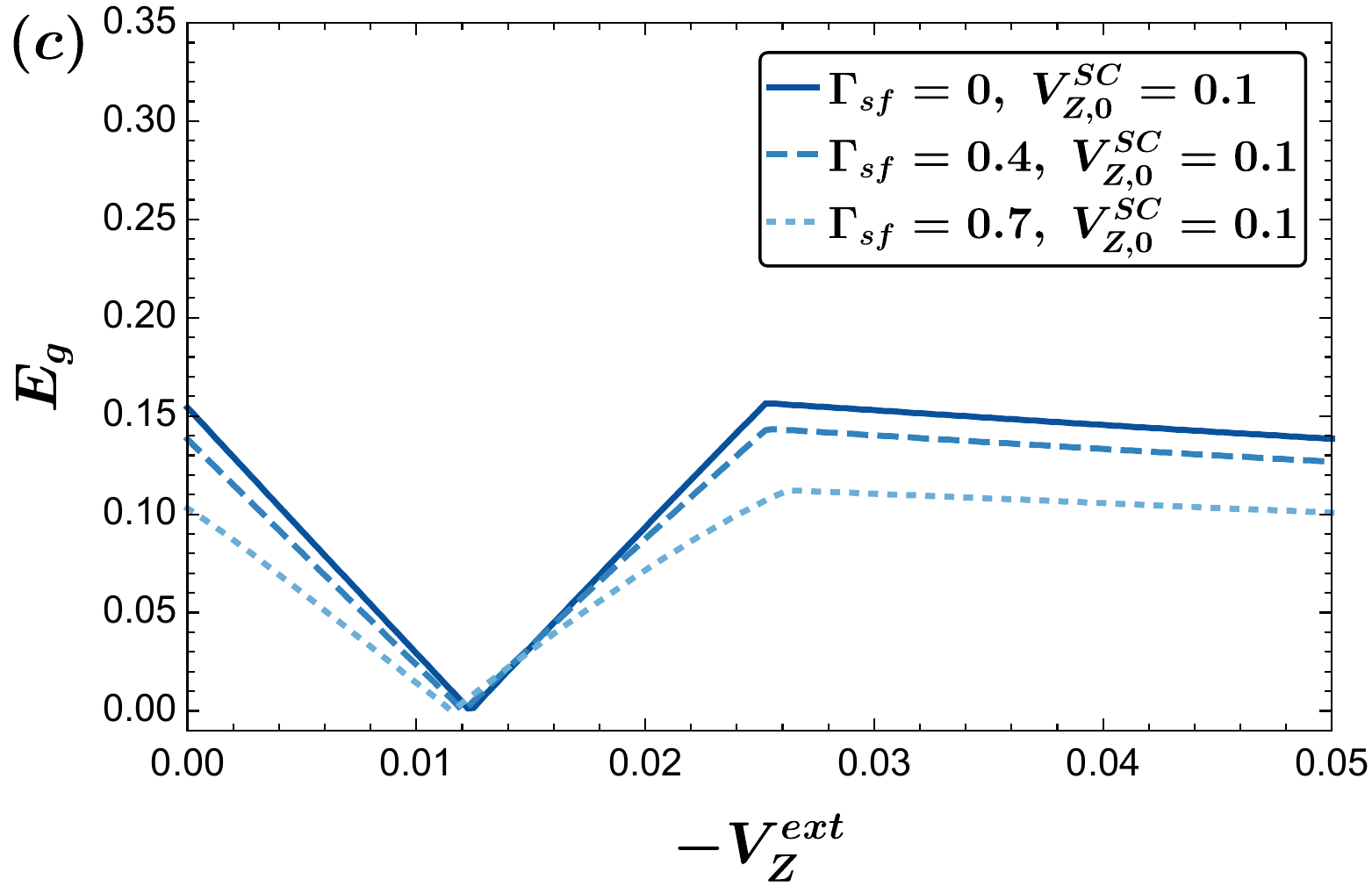}
	\caption{Energy gap in the proximitized nanowire as a function of external magnetic field $V_Z^{ext}=-\mu_B B/2$ for various values of SC-SM coupling $\gamma$, MI-induced Zeeman energy in the superconductor $V_{Z,0}^{SC}$, spin-orbit scattering $\Gamma_{so}$ and magnetic scattering $\Gamma_{sf}$. $\Gamma_{so}=0$ in (a),(c), $\Gamma_{sf}=0$ in (a),(b) and $\gamma=0.2$ in (b,c),(c). All curves are plotted for zero chemical potential in the nanowire. All energies are measured in units of the parent superconductor's bare gap $\Delta_{00}$.}
	\label{fig:Eg}
\end{figure*}

Figure~\ref{fig:phase_diag} shows the computed spectral gap as a function of the chemical potential $\mu$ in the nanowire and the external Zeeman field $V_Z^{ext}$ for $V_{Z,0}^{SC}=0.55$ in the absence of the magnetic and spin-orbit scattering in the SC.
The topological part of the phase diagram is marked red, solid black lines represent its boundary.
If the MI does not couple to the SC --- i.e., $V_{Z,0}^{SC} = 0$, --- the phase boundary is depicted by  dashed black lines.
In this case the orientation of the external field does not play any role which is exhibited by the symmetry of the two dashed curves with respect to the $V_Z^{ext}=0$ line.
However, if the MI induces Zeeman splitting in the SC, the relative orientation of the two fields --- $V_Z^{ext}$ and $V_{Z,0}^{SC}$ --- is important.
In the antiparallel configuration, which corresponds to the case of $V_Z^{ext} < 0$ in Fig.~\ref{fig:phase_diag}, the Zeeman splittings created by the MI and the applied field have opposite signs in the SC and the same sign in the SM.
As a result, the topological phase is achieved at smaller values of $|V_Z^{ext}|$ compared to the case when $V_{Z,0}^{SC} = 0$.
On the other hand, when the direction of $V_Z^{ext}$ is parallel to the magnetization of the MI --- $V_Z^{ext} > 0$ in Fig.~\ref{fig:phase_diag} --- the Zeeman splittings have the same sign in the SC but the opposite signs in the SM.
This is an unfavorable configuration for the topological phase: the Clogston limit in this case is reached at smaller values of the applied field, while the phase transition to the $p$-wave superconductivity requires a larger applied field compared to the case with no MI.

We now proceed to studying individual effects of the Zeeman splitting, spin-orbit and magnetic scattering on the topological gap.
To this end, we fix the chemical potential in the nanowire to zero and plot the dependence of the gap on $V_Z^{ext}$ in Fig.~\ref{fig:Eg} for the antiparallel configuration of the applied field and the MI magnetization, which is advantageous for the topological phase.
First, we consider impact of $V_{Z,0}^{SC}$ in Fig.~\ref{fig:Eg}(a) in the absence of the spin-orbit and magnetic scattering.
The black dotted curve in Fig.~\ref{fig:Eg}(a) shows the typical behavior of the energy gap as the system undergoes the topological transition between the $s$-wave (small $|V_Z^{ext}|$) and the $p$-wave (large $|V_Z^{ext}|$) superconductivity in the weak coupling regime ($\gamma=0.2$) in the absence of any zero-field Zeeman splitting in the SC.
The kink exhibited by the curve at $-V_Z^{ext}\approx 0.027$ appears where the minimum gap as a function of momentum jumps from $k=0$ to $k \sim k_F$.
Inducing finite Zeeman splitting in the SC results in the decrease of the critical Zeeman field as illustrated by the dashed line.
Analogous reduction of the critical field can be achieved by lowering the coupling between the SM and the SC (grey dotted line), but that also results in the reduction of the topological gap, whereas addition of the Zeeman energy to the SC keeps the gap approximately the same as long as $V_{Z,0}^{SC}$ is below the Clogston limit.
Adding Zeeman energy larger than the Clogston limit leads to an interesting behavior; see solid line in Fig.~\ref{fig:Eg}(a).
In this case at zero field the superconductivity in the parent SC is broken due to the proximitizing MI, but application of the external field restores it back.
As a result, the system undergoes the first order transition from the normal phase directly into the $p$-wave superconducting phase with the topological gap close to the one of the MI-free system.

As we have previously pointed out while discussing Fig.~\ref{fig:VZSM_VZSC}, spin-orbit scattering can quench the Zeeman effect in the superconductor, enhancing its Clogston limit, but also reducing the Zeeman energy effectively transferred to the nanowire.
This is illustrated in Fig.~\ref{fig:Eg}(b) where we plot the gap as a function of the external field for different values of $\Gamma_{so}$.
We consider the case when Zeeman energy above the Clogston limit is induced in the superconductor [solid line, same as in Fig.~\ref{fig:Eg}(a)].
Increasing the spin-orbit scattering in the parent SC restores its superconductivity, and a regular transition between the $s$- and $p$-wave phase is observed at finite $\Gamma_{so}$.
Further increase of $\Gamma_{so}$ leads to the enhancement of the critical field due to the suppression of the Zeeman splitting in the superconductor.

Effects of the magnetic impurity scattering on the topological gap are illustrated in Fig.~\ref{fig:Eg}(c).
As has been discussed in Section~\ref{sec:SC_calc}, magnetic scattering reduces the pairing potential in the parent SC.
This leads to a minute reduction of the critical field in Fig.~\ref{fig:Eg}(c) and a visible suppression of the topological gap.
Overall, we conclude that the presence of the magnetic scattering in the parent SC is not desirable for creating the topological phase in the nanowire.

\section{\label{sec:conclusion}conclusion}

In this paper we have presented a theoretical approach to calculate topological properties of nanowires proximitized by a bilayer of a disordered superconductor and a magnetic insulator.
We have taken into account the proximity effect between the superconductor and the magnetic insulator by means of the Usadel approach, and shown how different mechanisms (Zeeman splitting, magnetic and spin-orbit scattering) change the intrinsic properties of the superconductor.
We have then calculated how the bilayer changes the topological properties of a proximate semiconducting wire.
In particular, similar to other recent theoretical results \cite{Woods2020,Liu2020b,Escribano2020,Poyhonen2020,Langbehn2020} we report that a finite Zeeman energy in the semiconductor, either induced by coupling to a magnetic insulator or by an applied magnetic field, is required to enter the topological phase -- a bound we estimate analytically.
When introducing this Zeeman energy as a result of the applied magnetic field, we observe that the critical field and the topological gap depend on the relative orientation of the applied field and the MI-induced Zeeman splitting in the superconductor: only in the configuration when they are antiparallel one finds reduced critical field and larger stability with respect to the absolute magnitude of the applied field.
Introducing magnetic and spin-orbit scattering to the superconductor, we find that the former in general is detrimental to the topological phase -- it reduces the Clogston limit in the superconductor and leads to a smaller topological gap.
At the same time, spin-orbit scattering quenches the magnetic response of the superconductor, allowing it to sustain larger Zeeman fields, but also increases the critical field required to reach the topological phase.

In general, if magnetic insulators are used to decrease the critical magnetic field for entering the topological phase, when they proximate the superconductor, particular attention must be paid to the orientation of the magnetization and to the selection of a material that induces an exchange field that is neither too small nor too large to destroy the superconductivity.
Some degree of control of the induced field may be attained by considering the multi-domain MI structures and in fact the recent experiment \cite{Vaitiekenas2020} was performed in this regime. 
However, in practice the lack of control over the size of every domain renders a practical usage of this regime problematic.
The other unwanted effect of using magnetic insulator-superconductor bilayers is the magnetic impurity scattering. In order to reduce it, likely the high quality of the interface between the magnetic insulator and the superconductor is paramount.
One other possible axis of optimization is addition of the spin-orbit scattering to the superconductor -- it enhances stability to the applied magnetic fields while still providing the benefit of overall-smaller operating fields afforded by some amount of zero-field spin splitting in the superconductor.

Current progress in constructing nanowires that produce topological excitations leads to an increasing complexity of devices, including the use of engineered stacks consisting of various materials.
Beyond magnetic insulators, other materials that could be used in such stacks include larger-gap superconductors, e.g., Pb \cite{Kanne2020}, to increase the pairing potential of the whole stack, or a normal metal with large spin-orbit coupling, which could add spin-orbit scattering to the superconductor (thereby increasing the Clogston limit).
An approach presented here based on the Usadel equation treatment of the superconducting stacks and the BdG treatment of the semiconductor can be generalized to study topological properties of heterostructures containing these (and other) materials.
In addition, this approach is scalable to three dimensions and can account for the effects of inhomogeneities, multiple magnetic domains, etc.
Thus it should be instrumental in the quest for finding an optimal Majorana platform.

\begin{acknowledgments}

We thank Bela Bauer, Roman Lutchyn, Saulius Vaitiek\.enas, Charles M. Marcus, Chun-Xiao Liu, Michael Wimmer and Chetan Nayak for useful discussions. AK also thanks Matthew P. A. Fisher and Andrea F. Young for valuable comments. PAL acknowledges support from the NSF C-Accel Track C grant 2040620. J.A.'s work was supported by Army Research Office under Grant Award W911NF17- 1-0323; the National Science Foundation through grant DMR-1723367; the Caltech Institute for Quantum Information and Matter, an NSF Physics Frontiers Center with support of the Gordon and Betty Moore Foundation through Grant GBMF1250; and the Walter Burke Institute for Theoretical Physics at Caltech. The final stage of this work was in part based on support by the U.S. Department of Energy, Office of Science through the Quantum Science Center (QSC), a National Quantum Information Science Research Center. AK's work was supported by Microsoft corporation. Use was made of computational facilities purchased with funds from the National Science Foundation (CNS-1725797) and administered by the Center for Scientific Computing (CSC). The CSC is supported by the California NanoSystems Institute and the Materials Research Science and Engineering Center (MRSEC; NSF DMR 1720256) at UC Santa Barbara.

\end{acknowledgments}

\bibliography{FMI-SC-SM}

\begin{thebibliography}{60}%
\makeatletter
\providecommand \@ifxundefined [1]{%
 \@ifx{#1\undefined}
}%
\providecommand \@ifnum [1]{%
 \ifnum #1\expandafter \@firstoftwo
 \else \expandafter \@secondoftwo
 \fi
}%
\providecommand \@ifx [1]{%
 \ifx #1\expandafter \@firstoftwo
 \else \expandafter \@secondoftwo
 \fi
}%
\providecommand \natexlab [1]{#1}%
\providecommand \enquote  [1]{``#1''}%
\providecommand \bibnamefont  [1]{#1}%
\providecommand \bibfnamefont [1]{#1}%
\providecommand \citenamefont [1]{#1}%
\providecommand \href@noop [0]{\@secondoftwo}%
\providecommand \href [0]{\begingroup \@sanitize@url \@href}%
\providecommand \@href[1]{\@@startlink{#1}\@@href}%
\providecommand \@@href[1]{\endgroup#1\@@endlink}%
\providecommand \@sanitize@url [0]{\catcode `\\12\catcode `\$12\catcode
  `\&12\catcode `\#12\catcode `\^12\catcode `\_12\catcode `\%12\relax}%
\providecommand \@@startlink[1]{}%
\providecommand \@@endlink[0]{}%
\providecommand \url  [0]{\begingroup\@sanitize@url \@url }%
\providecommand \@url [1]{\endgroup\@href {#1}{\urlprefix }}%
\providecommand \urlprefix  [0]{URL }%
\providecommand \Eprint [0]{\href }%
\providecommand \doibase [0]{http://dx.doi.org/}%
\providecommand \selectlanguage [0]{\@gobble}%
\providecommand \bibinfo  [0]{\@secondoftwo}%
\providecommand \bibfield  [0]{\@secondoftwo}%
\providecommand \translation [1]{[#1]}%
\providecommand \BibitemOpen [0]{}%
\providecommand \bibitemStop [0]{}%
\providecommand \bibitemNoStop [0]{.\EOS\space}%
\providecommand \EOS [0]{\spacefactor3000\relax}%
\providecommand \BibitemShut  [1]{\csname bibitem#1\endcsname}%
\let\auto@bib@innerbib\@empty
\bibitem [{\citenamefont {Freedman}\ \emph {et~al.}(2003)\citenamefont
  {Freedman}, \citenamefont {Kitaev}, \citenamefont {Larsen},\ and\
  \citenamefont {Wang}}]{Freedman2003}%
  \BibitemOpen
  \bibfield  {author} {\bibinfo {author} {\bibfnamefont {M.~H.}\ \bibnamefont
  {Freedman}}, \bibinfo {author} {\bibfnamefont {A.}~\bibnamefont {Kitaev}},
  \bibinfo {author} {\bibfnamefont {M.~J.}\ \bibnamefont {Larsen}}, \ and\
  \bibinfo {author} {\bibfnamefont {Z.}~\bibnamefont {Wang}},\ }\href {\doibase
  10.1090/S0273-0979-02-00964-3} {\bibfield  {journal} {\bibinfo  {journal}
  {Bull. Am. Math. Soc.}\ }\textbf {\bibinfo {volume} {40}},\ \bibinfo {pages}
  {31} (\bibinfo {year} {2003})}\BibitemShut {NoStop}%
\bibitem [{\citenamefont {Nayak}\ \emph {et~al.}(2008)\citenamefont {Nayak},
  \citenamefont {Simon}, \citenamefont {Stern}, \citenamefont {Freedman},\ and\
  \citenamefont {{Das Sarma}}}]{Nayak2008}%
  \BibitemOpen
  \bibfield  {author} {\bibinfo {author} {\bibfnamefont {C.}~\bibnamefont
  {Nayak}}, \bibinfo {author} {\bibfnamefont {S.~H.}\ \bibnamefont {Simon}},
  \bibinfo {author} {\bibfnamefont {A.}~\bibnamefont {Stern}}, \bibinfo
  {author} {\bibfnamefont {M.}~\bibnamefont {Freedman}}, \ and\ \bibinfo
  {author} {\bibfnamefont {S.}~\bibnamefont {{Das Sarma}}},\ }\href {\doibase
  10.1103/RevModPhys.80.1083} {\bibfield  {journal} {\bibinfo  {journal} {Rev.
  Mod. Phys.}\ }\textbf {\bibinfo {volume} {80}},\ \bibinfo {pages} {1083}
  (\bibinfo {year} {2008})},\ \Eprint {http://arxiv.org/abs/0707.1889}
  {arXiv:0707.1889} \BibitemShut {NoStop}%
\bibitem [{\citenamefont {Sarma}\ \emph {et~al.}(2015)\citenamefont {Sarma},
  \citenamefont {Freedman},\ and\ \citenamefont {Nayak}}]{Dassarma2015}%
  \BibitemOpen
  \bibfield  {author} {\bibinfo {author} {\bibfnamefont {S.~D.}\ \bibnamefont
  {Sarma}}, \bibinfo {author} {\bibfnamefont {M.}~\bibnamefont {Freedman}}, \
  and\ \bibinfo {author} {\bibfnamefont {C.}~\bibnamefont {Nayak}},\ }\href
  {\doibase 10.1038/npjqi.2015.1} {\bibfield  {journal} {\bibinfo  {journal}
  {npj Quantum Inf.}\ }\textbf {\bibinfo {volume} {1}},\ \bibinfo {pages}
  {15001} (\bibinfo {year} {2015})},\ \Eprint {http://arxiv.org/abs/1501.02813}
  {arXiv:1501.02813} \BibitemShut {NoStop}%
\bibitem [{\citenamefont {Alicea}(2010)}]{Alicea2010}%
  \BibitemOpen
  \bibfield  {author} {\bibinfo {author} {\bibfnamefont {J.}~\bibnamefont
  {Alicea}},\ }\href {\doibase 10.1103/PhysRevB.81.125318} {\bibfield
  {journal} {\bibinfo  {journal} {Phys. Rev. B}\ }\textbf {\bibinfo {volume}
  {81}},\ \bibinfo {pages} {125318} (\bibinfo {year} {2010})},\ \Eprint
  {http://arxiv.org/abs/0912.2115} {arXiv:0912.2115} \BibitemShut {NoStop}%
\bibitem [{\citenamefont {Lutchyn}\ \emph {et~al.}(2010)\citenamefont
  {Lutchyn}, \citenamefont {Sau},\ and\ \citenamefont {{Das
  Sarma}}}]{Lutchyn2010}%
  \BibitemOpen
  \bibfield  {author} {\bibinfo {author} {\bibfnamefont {R.~M.}\ \bibnamefont
  {Lutchyn}}, \bibinfo {author} {\bibfnamefont {J.~D.}\ \bibnamefont {Sau}}, \
  and\ \bibinfo {author} {\bibfnamefont {S.}~\bibnamefont {{Das Sarma}}},\
  }\href {\doibase 10.1103/PhysRevLett.105.077001} {\bibfield  {journal}
  {\bibinfo  {journal} {Phys. Rev. Lett.}\ }\textbf {\bibinfo {volume} {105}},\
  \bibinfo {pages} {077001} (\bibinfo {year} {2010})},\ \Eprint
  {http://arxiv.org/abs/1002.4033} {arXiv:1002.4033} \BibitemShut {NoStop}%
\bibitem [{\citenamefont {Oreg}\ \emph {et~al.}(2010)\citenamefont {Oreg},
  \citenamefont {Refael},\ and\ \citenamefont {{Von Oppen}}}]{Oreg2010}%
  \BibitemOpen
  \bibfield  {author} {\bibinfo {author} {\bibfnamefont {Y.}~\bibnamefont
  {Oreg}}, \bibinfo {author} {\bibfnamefont {G.}~\bibnamefont {Refael}}, \ and\
  \bibinfo {author} {\bibfnamefont {F.}~\bibnamefont {{Von Oppen}}},\ }\href
  {\doibase 10.1103/PhysRevLett.105.177002} {\bibfield  {journal} {\bibinfo
  {journal} {Phys. Rev. Lett.}\ }\textbf {\bibinfo {volume} {105}},\ \bibinfo
  {pages} {177002} (\bibinfo {year} {2010})},\ \Eprint
  {http://arxiv.org/abs/1003.1145} {arXiv:1003.1145} \BibitemShut {NoStop}%
\bibitem [{\citenamefont {Lutchyn}\ \emph {et~al.}(2018)\citenamefont
  {Lutchyn}, \citenamefont {Bakkers}, \citenamefont {Kouwenhoven},
  \citenamefont {Krogstrup}, \citenamefont {Marcus},\ and\ \citenamefont
  {Oreg}}]{Lutchyn2018}%
  \BibitemOpen
  \bibfield  {author} {\bibinfo {author} {\bibfnamefont {R.~M.}\ \bibnamefont
  {Lutchyn}}, \bibinfo {author} {\bibfnamefont {E.~P. A.~M.}\ \bibnamefont
  {Bakkers}}, \bibinfo {author} {\bibfnamefont {L.~P.}\ \bibnamefont
  {Kouwenhoven}}, \bibinfo {author} {\bibfnamefont {P.}~\bibnamefont
  {Krogstrup}}, \bibinfo {author} {\bibfnamefont {C.~M.}\ \bibnamefont
  {Marcus}}, \ and\ \bibinfo {author} {\bibfnamefont {Y.}~\bibnamefont
  {Oreg}},\ }\href {\doibase 10.1038/s41578-018-0003-1} {\bibfield  {journal}
  {\bibinfo  {journal} {Nat. Rev. Mater.}\ }\textbf {\bibinfo {volume} {3}},\
  \bibinfo {pages} {52} (\bibinfo {year} {2018})},\ \Eprint
  {http://arxiv.org/abs/1707.04899v2} {arXiv:1707.04899v2} \BibitemShut
  {NoStop}%
\bibitem [{\citenamefont {Antipov}\ \emph {et~al.}(2018)\citenamefont
  {Antipov}, \citenamefont {Bargerbos}, \citenamefont {Winkler}, \citenamefont
  {Bauer}, \citenamefont {Rossi},\ and\ \citenamefont {Lutchyn}}]{Antipov2018}%
  \BibitemOpen
  \bibfield  {author} {\bibinfo {author} {\bibfnamefont {A.~E.}\ \bibnamefont
  {Antipov}}, \bibinfo {author} {\bibfnamefont {A.}~\bibnamefont {Bargerbos}},
  \bibinfo {author} {\bibfnamefont {G.~W.}\ \bibnamefont {Winkler}}, \bibinfo
  {author} {\bibfnamefont {B.}~\bibnamefont {Bauer}}, \bibinfo {author}
  {\bibfnamefont {E.}~\bibnamefont {Rossi}}, \ and\ \bibinfo {author}
  {\bibfnamefont {R.~M.}\ \bibnamefont {Lutchyn}},\ }\href {\doibase
  10.1103/PhysRevX.8.031041} {\bibfield  {journal} {\bibinfo  {journal} {Phys.
  Rev. X}\ }\textbf {\bibinfo {volume} {8}},\ \bibinfo {pages} {31041}
  (\bibinfo {year} {2018})},\ \Eprint {http://arxiv.org/abs/1801.02616}
  {arXiv:1801.02616} \BibitemShut {NoStop}%
\bibitem [{\citenamefont {Lutchyn}\ \emph {et~al.}(2012)\citenamefont
  {Lutchyn}, \citenamefont {Stanescu},\ and\ \citenamefont {{Das
  Sarma}}}]{Lutchyn2012}%
  \BibitemOpen
  \bibfield  {author} {\bibinfo {author} {\bibfnamefont {R.~M.}\ \bibnamefont
  {Lutchyn}}, \bibinfo {author} {\bibfnamefont {T.~D.}\ \bibnamefont
  {Stanescu}}, \ and\ \bibinfo {author} {\bibfnamefont {S.}~\bibnamefont {{Das
  Sarma}}},\ }\href {\doibase 10.1103/PhysRevB.85.140513} {\bibfield  {journal}
  {\bibinfo  {journal} {Phys. Rev. B - Condens. Matter Mater. Phys.}\ }\textbf
  {\bibinfo {volume} {85}},\ \bibinfo {pages} {140513} (\bibinfo {year}
  {2012})},\ \Eprint {http://arxiv.org/abs/1110.5643} {arXiv:1110.5643}
  \BibitemShut {NoStop}%
\bibitem [{\citenamefont {Cole}\ \emph {et~al.}(2016)\citenamefont {Cole},
  \citenamefont {Sau},\ and\ \citenamefont {{Das Sarma}}}]{Cole2016}%
  \BibitemOpen
  \bibfield  {author} {\bibinfo {author} {\bibfnamefont {W.~S.}\ \bibnamefont
  {Cole}}, \bibinfo {author} {\bibfnamefont {J.~D.}\ \bibnamefont {Sau}}, \
  and\ \bibinfo {author} {\bibfnamefont {S.}~\bibnamefont {{Das Sarma}}},\
  }\href {\doibase 10.1103/PhysRevB.94.140505} {\bibfield  {journal} {\bibinfo
  {journal} {Phys. Rev. B}\ }\textbf {\bibinfo {volume} {94}} (\bibinfo {year}
  {2016}),\ 10.1103/PhysRevB.94.140505},\ \Eprint
  {http://arxiv.org/abs/1603.03780} {arXiv:1603.03780} \BibitemShut {NoStop}%
\bibitem [{\citenamefont {Nijholt}\ and\ \citenamefont
  {Akhmerov}(2016)}]{Nijholt2016}%
  \BibitemOpen
  \bibfield  {author} {\bibinfo {author} {\bibfnamefont {B.}~\bibnamefont
  {Nijholt}}\ and\ \bibinfo {author} {\bibfnamefont {A.~R.}\ \bibnamefont
  {Akhmerov}},\ }\href {\doibase 10.1103/PhysRevB.93.235434} {\bibfield
  {journal} {\bibinfo  {journal} {Phys. Rev. B}\ }\textbf {\bibinfo {volume}
  {93}},\ \bibinfo {pages} {235434} (\bibinfo {year} {2016})},\ \Eprint
  {http://arxiv.org/abs/1509.02675} {arXiv:1509.02675} \BibitemShut {NoStop}%
\bibitem [{\citenamefont {Winkler}\ \emph {et~al.}(2019)\citenamefont
  {Winkler}, \citenamefont {Antipov}, \citenamefont {{Van Heck}}, \citenamefont
  {Soluyanov}, \citenamefont {Glazman}, \citenamefont {Wimmer},\ and\
  \citenamefont {Lutchyn}}]{Winkler2019}%
  \BibitemOpen
  \bibfield  {author} {\bibinfo {author} {\bibfnamefont {G.~W.}\ \bibnamefont
  {Winkler}}, \bibinfo {author} {\bibfnamefont {A.~E.}\ \bibnamefont
  {Antipov}}, \bibinfo {author} {\bibfnamefont {B.}~\bibnamefont {{Van Heck}}},
  \bibinfo {author} {\bibfnamefont {A.~A.}\ \bibnamefont {Soluyanov}}, \bibinfo
  {author} {\bibfnamefont {L.~I.}\ \bibnamefont {Glazman}}, \bibinfo {author}
  {\bibfnamefont {M.}~\bibnamefont {Wimmer}}, \ and\ \bibinfo {author}
  {\bibfnamefont {R.~M.}\ \bibnamefont {Lutchyn}},\ }\href {\doibase
  10.1103/PhysRevB.99.245408} {\bibfield  {journal} {\bibinfo  {journal} {Phys.
  Rev. B}\ }\textbf {\bibinfo {volume} {99}},\ \bibinfo {pages} {245408}
  (\bibinfo {year} {2019})}\BibitemShut {NoStop}%
\bibitem [{\citenamefont {Karzig}\ \emph {et~al.}(2017)\citenamefont {Karzig},
  \citenamefont {Knapp}, \citenamefont {Lutchyn}, \citenamefont {Bonderson},
  \citenamefont {Hastings}, \citenamefont {Nayak}, \citenamefont {Alicea},
  \citenamefont {Flensberg}, \citenamefont {Plugge}, \citenamefont {Oreg},
  \citenamefont {Marcus},\ and\ \citenamefont {Freedman}}]{Karzig2017}%
  \BibitemOpen
  \bibfield  {author} {\bibinfo {author} {\bibfnamefont {T.}~\bibnamefont
  {Karzig}}, \bibinfo {author} {\bibfnamefont {C.}~\bibnamefont {Knapp}},
  \bibinfo {author} {\bibfnamefont {R.~M.}\ \bibnamefont {Lutchyn}}, \bibinfo
  {author} {\bibfnamefont {P.}~\bibnamefont {Bonderson}}, \bibinfo {author}
  {\bibfnamefont {M.~B.}\ \bibnamefont {Hastings}}, \bibinfo {author}
  {\bibfnamefont {C.}~\bibnamefont {Nayak}}, \bibinfo {author} {\bibfnamefont
  {J.}~\bibnamefont {Alicea}}, \bibinfo {author} {\bibfnamefont
  {K.}~\bibnamefont {Flensberg}}, \bibinfo {author} {\bibfnamefont
  {S.}~\bibnamefont {Plugge}}, \bibinfo {author} {\bibfnamefont
  {Y.}~\bibnamefont {Oreg}}, \bibinfo {author} {\bibfnamefont {C.~M.}\
  \bibnamefont {Marcus}}, \ and\ \bibinfo {author} {\bibfnamefont {M.~H.}\
  \bibnamefont {Freedman}},\ }\href {\doibase 10.1103/PhysRevB.95.235305}
  {\bibfield  {journal} {\bibinfo  {journal} {Phys. Rev. B}\ }\textbf {\bibinfo
  {volume} {95}},\ \bibinfo {pages} {235305} (\bibinfo {year} {2017})},\
  \Eprint {http://arxiv.org/abs/1610.05289} {arXiv:1610.05289} \BibitemShut
  {NoStop}%
\bibitem [{\citenamefont {Vaitiekėnas}\ \emph {et~al.}(2020)\citenamefont
  {Vaitiekėnas}, \citenamefont {Liu}, \citenamefont {Krogstrup},\ and\
  \citenamefont {Marcus}}]{Vaitiekenas2020}%
  \BibitemOpen
  \bibfield  {author} {\bibinfo {author} {\bibfnamefont {S.}~\bibnamefont
  {Vaitiekėnas}}, \bibinfo {author} {\bibfnamefont {Y.}~\bibnamefont {Liu}},
  \bibinfo {author} {\bibfnamefont {P.}~\bibnamefont {Krogstrup}}, \ and\
  \bibinfo {author} {\bibfnamefont {C.~M.}\ \bibnamefont {Marcus}},\ }\href
  {\doibase 10.1038/s41567-020-1017-3} {\bibfield  {journal} {\bibinfo
  {journal} {Nat. Phys.}\ } (\bibinfo {year} {2020}),\
  10.1038/s41567-020-1017-3}\BibitemShut {NoStop}%
\bibitem [{\citenamefont {Liu}\ \emph {et~al.}(2020{\natexlab{a}})\citenamefont
  {Liu}, \citenamefont {Vaitiekėnas}, \citenamefont
  {Mart{\'{i}}-S{\'{a}}nchez}, \citenamefont {Koch}, \citenamefont {Hart},
  \citenamefont {Cui}, \citenamefont {Kanne}, \citenamefont {Khan},
  \citenamefont {Tanta}, \citenamefont {Upadhyay}, \citenamefont {Cachaza},
  \citenamefont {Marcus}, \citenamefont {Arbiol}, \citenamefont {Moler},\ and\
  \citenamefont {Krogstrup}}]{Liu2020a}%
  \BibitemOpen
  \bibfield  {author} {\bibinfo {author} {\bibfnamefont {Y.}~\bibnamefont
  {Liu}}, \bibinfo {author} {\bibfnamefont {S.}~\bibnamefont {Vaitiekėnas}},
  \bibinfo {author} {\bibfnamefont {S.}~\bibnamefont
  {Mart{\'{i}}-S{\'{a}}nchez}}, \bibinfo {author} {\bibfnamefont
  {C.}~\bibnamefont {Koch}}, \bibinfo {author} {\bibfnamefont {S.}~\bibnamefont
  {Hart}}, \bibinfo {author} {\bibfnamefont {Z.}~\bibnamefont {Cui}}, \bibinfo
  {author} {\bibfnamefont {T.}~\bibnamefont {Kanne}}, \bibinfo {author}
  {\bibfnamefont {S.~A.}\ \bibnamefont {Khan}}, \bibinfo {author}
  {\bibfnamefont {R.}~\bibnamefont {Tanta}}, \bibinfo {author} {\bibfnamefont
  {S.}~\bibnamefont {Upadhyay}}, \bibinfo {author} {\bibfnamefont {M.~E.}\
  \bibnamefont {Cachaza}}, \bibinfo {author} {\bibfnamefont {C.~M.}\
  \bibnamefont {Marcus}}, \bibinfo {author} {\bibfnamefont {J.}~\bibnamefont
  {Arbiol}}, \bibinfo {author} {\bibfnamefont {K.~A.}\ \bibnamefont {Moler}}, \
  and\ \bibinfo {author} {\bibfnamefont {P.}~\bibnamefont {Krogstrup}},\ }\href
  {\doibase 10.1021/acs.nanolett.9b04187} {\bibfield  {journal} {\bibinfo
  {journal} {Nano Lett.}\ }\textbf {\bibinfo {volume} {20}},\ \bibinfo {pages}
  {456} (\bibinfo {year} {2020}{\natexlab{a}})},\ \Eprint
  {http://arxiv.org/abs/1910.03364} {arXiv:1910.03364} \BibitemShut {NoStop}%
\bibitem [{\citenamefont {Tedrow}\ \emph {et~al.}(1986)\citenamefont {Tedrow},
  \citenamefont {Tkaczyk},\ and\ \citenamefont {Kumar}}]{Tedrow1986}%
  \BibitemOpen
  \bibfield  {author} {\bibinfo {author} {\bibfnamefont {P.~M.}\ \bibnamefont
  {Tedrow}}, \bibinfo {author} {\bibfnamefont {J.~E.}\ \bibnamefont {Tkaczyk}},
  \ and\ \bibinfo {author} {\bibfnamefont {A.}~\bibnamefont {Kumar}},\ }\href
  {\doibase 10.1103/PhysRevLett.56.1746} {\bibfield  {journal} {\bibinfo
  {journal} {Phys. Rev. Lett.}\ }\textbf {\bibinfo {volume} {56}},\ \bibinfo
  {pages} {1746} (\bibinfo {year} {1986})}\BibitemShut {NoStop}%
\bibitem [{\citenamefont {Hao}\ \emph {et~al.}(1991)\citenamefont {Hao},
  \citenamefont {Moodera},\ and\ \citenamefont {Meservey}}]{Hao1991}%
  \BibitemOpen
  \bibfield  {author} {\bibinfo {author} {\bibfnamefont {X.}~\bibnamefont
  {Hao}}, \bibinfo {author} {\bibfnamefont {J.~S.}\ \bibnamefont {Moodera}}, \
  and\ \bibinfo {author} {\bibfnamefont {R.}~\bibnamefont {Meservey}},\ }\href
  {\doibase 10.1103/PhysRevLett.67.1342} {\bibfield  {journal} {\bibinfo
  {journal} {Phys. Rev. Lett.}\ }\textbf {\bibinfo {volume} {67}},\ \bibinfo
  {pages} {1342} (\bibinfo {year} {1991})}\BibitemShut {NoStop}%
\bibitem [{\citenamefont {Wolf}\ \emph {et~al.}(2014)\citenamefont {Wolf},
  \citenamefont {S{\"{u}}rgers}, \citenamefont {Fischer},\ and\ \citenamefont
  {Beckmann}}]{Wolf2014}%
  \BibitemOpen
  \bibfield  {author} {\bibinfo {author} {\bibfnamefont {M.~J.}\ \bibnamefont
  {Wolf}}, \bibinfo {author} {\bibfnamefont {C.}~\bibnamefont {S{\"{u}}rgers}},
  \bibinfo {author} {\bibfnamefont {G.}~\bibnamefont {Fischer}}, \ and\
  \bibinfo {author} {\bibfnamefont {D.}~\bibnamefont {Beckmann}},\ }\href
  {\doibase 10.1103/PhysRevB.90.144509} {\bibfield  {journal} {\bibinfo
  {journal} {Phys. Rev. B}\ }\textbf {\bibinfo {volume} {90}},\ \bibinfo
  {pages} {144509} (\bibinfo {year} {2014})},\ \Eprint
  {http://arxiv.org/abs/1408.0176} {arXiv:1408.0176} \BibitemShut {NoStop}%
\bibitem [{\citenamefont {Strambini}\ \emph {et~al.}(2017)\citenamefont
  {Strambini}, \citenamefont {Golovach}, \citenamefont {{De Simoni}},
  \citenamefont {Moodera}, \citenamefont {Bergeret},\ and\ \citenamefont
  {Giazotto}}]{Strambini2017}%
  \BibitemOpen
  \bibfield  {author} {\bibinfo {author} {\bibfnamefont {E.}~\bibnamefont
  {Strambini}}, \bibinfo {author} {\bibfnamefont {V.~N.}\ \bibnamefont
  {Golovach}}, \bibinfo {author} {\bibfnamefont {G.}~\bibnamefont {{De
  Simoni}}}, \bibinfo {author} {\bibfnamefont {J.~S.}\ \bibnamefont {Moodera}},
  \bibinfo {author} {\bibfnamefont {F.~S.}\ \bibnamefont {Bergeret}}, \ and\
  \bibinfo {author} {\bibfnamefont {F.}~\bibnamefont {Giazotto}},\ }\href
  {\doibase 10.1103/PhysRevMaterials.1.054402} {\bibfield  {journal} {\bibinfo
  {journal} {Phys. Rev. Mater.}\ }\textbf {\bibinfo {volume} {1}},\ \bibinfo
  {pages} {54402} (\bibinfo {year} {2017})},\ \Eprint
  {http://arxiv.org/abs/1705.04795} {arXiv:1705.04795} \BibitemShut {NoStop}%
\bibitem [{\citenamefont {Hijano}\ \emph {et~al.}(2020)\citenamefont {Hijano},
  \citenamefont {Ili{\'{c}}}, \citenamefont {Rouco}, \citenamefont {Orellana},
  \citenamefont {Ilyn}, \citenamefont {Rogero}, \citenamefont {Virtanen},
  \citenamefont {Heikkil{\"{a}}}, \citenamefont {Khorshidian}, \citenamefont
  {Spies}, \citenamefont {Giazotto}, \citenamefont {Strambini},\ and\
  \citenamefont {Bergeret}}]{Hijano2021}%
  \BibitemOpen
  \bibfield  {author} {\bibinfo {author} {\bibfnamefont {A.}~\bibnamefont
  {Hijano}}, \bibinfo {author} {\bibfnamefont {S.}~\bibnamefont {Ili{\'{c}}}},
  \bibinfo {author} {\bibfnamefont {M.}~\bibnamefont {Rouco}}, \bibinfo
  {author} {\bibfnamefont {C.~G.}\ \bibnamefont {Orellana}}, \bibinfo {author}
  {\bibfnamefont {M.}~\bibnamefont {Ilyn}}, \bibinfo {author} {\bibfnamefont
  {C.}~\bibnamefont {Rogero}}, \bibinfo {author} {\bibfnamefont
  {P.}~\bibnamefont {Virtanen}}, \bibinfo {author} {\bibfnamefont {T.~T.}\
  \bibnamefont {Heikkil{\"{a}}}}, \bibinfo {author} {\bibfnamefont
  {S.}~\bibnamefont {Khorshidian}}, \bibinfo {author} {\bibfnamefont
  {M.}~\bibnamefont {Spies}}, \bibinfo {author} {\bibfnamefont
  {F.}~\bibnamefont {Giazotto}}, \bibinfo {author} {\bibfnamefont
  {E.}~\bibnamefont {Strambini}}, \ and\ \bibinfo {author} {\bibfnamefont
  {F.~S.}\ \bibnamefont {Bergeret}},\ }\href {https://arxiv.org/abs/2012.15549}
  {\  (\bibinfo {year} {2020})},\ \Eprint {http://arxiv.org/abs/2012.15549}
  {arXiv:2012.15549} \BibitemShut {NoStop}%
\bibitem [{\citenamefont {Liu}\ \emph {et~al.}(2020{\natexlab{b}})\citenamefont
  {Liu}, \citenamefont {Luchini}, \citenamefont {Mart{\'{i}}-S{\'{a}}nchez},
  \citenamefont {Koch}, \citenamefont {Schuwalow}, \citenamefont {Khan},
  \citenamefont {Stankevi{\v{c}}}, \citenamefont {Francoual}, \citenamefont
  {Mardegan}, \citenamefont {Krieger}, \citenamefont {Strocov}, \citenamefont
  {Stahn}, \citenamefont {Vaz}, \citenamefont {Ramakrishnan}, \citenamefont
  {Staub}, \citenamefont {Lefmann}, \citenamefont {Aeppli}, \citenamefont
  {Arbiol},\ and\ \citenamefont {Krogstrup}}]{Liu2020}%
  \BibitemOpen
  \bibfield  {author} {\bibinfo {author} {\bibfnamefont {Y.}~\bibnamefont
  {Liu}}, \bibinfo {author} {\bibfnamefont {A.}~\bibnamefont {Luchini}},
  \bibinfo {author} {\bibfnamefont {S.}~\bibnamefont
  {Mart{\'{i}}-S{\'{a}}nchez}}, \bibinfo {author} {\bibfnamefont
  {C.}~\bibnamefont {Koch}}, \bibinfo {author} {\bibfnamefont {S.}~\bibnamefont
  {Schuwalow}}, \bibinfo {author} {\bibfnamefont {S.~A.}\ \bibnamefont {Khan}},
  \bibinfo {author} {\bibfnamefont {T.}~\bibnamefont {Stankevi{\v{c}}}},
  \bibinfo {author} {\bibfnamefont {S.}~\bibnamefont {Francoual}}, \bibinfo
  {author} {\bibfnamefont {J.~R.}\ \bibnamefont {Mardegan}}, \bibinfo {author}
  {\bibfnamefont {J.~A.}\ \bibnamefont {Krieger}}, \bibinfo {author}
  {\bibfnamefont {V.~N.}\ \bibnamefont {Strocov}}, \bibinfo {author}
  {\bibfnamefont {J.}~\bibnamefont {Stahn}}, \bibinfo {author} {\bibfnamefont
  {C.~A.}\ \bibnamefont {Vaz}}, \bibinfo {author} {\bibfnamefont
  {M.}~\bibnamefont {Ramakrishnan}}, \bibinfo {author} {\bibfnamefont
  {U.}~\bibnamefont {Staub}}, \bibinfo {author} {\bibfnamefont
  {K.}~\bibnamefont {Lefmann}}, \bibinfo {author} {\bibfnamefont
  {G.}~\bibnamefont {Aeppli}}, \bibinfo {author} {\bibfnamefont
  {J.}~\bibnamefont {Arbiol}}, \ and\ \bibinfo {author} {\bibfnamefont
  {P.}~\bibnamefont {Krogstrup}},\ }\href {\doibase 10.1021/acsami.9b15034}
  {\bibfield  {journal} {\bibinfo  {journal} {ACS Appl. Mater. Interfaces}\
  }\textbf {\bibinfo {volume} {12}},\ \bibinfo {pages} {8780} (\bibinfo {year}
  {2020}{\natexlab{b}})},\ \Eprint {http://arxiv.org/abs/1908.07096}
  {arXiv:1908.07096} \BibitemShut {NoStop}%
\bibitem [{\citenamefont {Woods}\ and\ \citenamefont
  {Stanescu}(2020)}]{Woods2020}%
  \BibitemOpen
  \bibfield  {author} {\bibinfo {author} {\bibfnamefont {B.~D.}\ \bibnamefont
  {Woods}}\ and\ \bibinfo {author} {\bibfnamefont {T.~D.}\ \bibnamefont
  {Stanescu}},\ }\href {http://arxiv.org/abs/2011.01933} {\  (\bibinfo {year}
  {2020})},\ \Eprint {http://arxiv.org/abs/2011.01933} {arXiv:2011.01933}
  \BibitemShut {NoStop}%
\bibitem [{\citenamefont {Liu}\ \emph {et~al.}(2020{\natexlab{c}})\citenamefont
  {Liu}, \citenamefont {Schuwalow}, \citenamefont {Liu}, \citenamefont
  {Vilkelis}, \citenamefont {Manesco}, \citenamefont {Krogstrup},\ and\
  \citenamefont {Wimmer}}]{Liu2020b}%
  \BibitemOpen
  \bibfield  {author} {\bibinfo {author} {\bibfnamefont {C.-X.}\ \bibnamefont
  {Liu}}, \bibinfo {author} {\bibfnamefont {S.}~\bibnamefont {Schuwalow}},
  \bibinfo {author} {\bibfnamefont {Y.}~\bibnamefont {Liu}}, \bibinfo {author}
  {\bibfnamefont {K.}~\bibnamefont {Vilkelis}}, \bibinfo {author}
  {\bibfnamefont {A.~L.~R.}\ \bibnamefont {Manesco}}, \bibinfo {author}
  {\bibfnamefont {P.}~\bibnamefont {Krogstrup}}, \ and\ \bibinfo {author}
  {\bibfnamefont {M.}~\bibnamefont {Wimmer}},\ }\href
  {http://arxiv.org/abs/2011.06567} {\  (\bibinfo {year}
  {2020}{\natexlab{c}})},\ \Eprint {http://arxiv.org/abs/2011.06567}
  {arXiv:2011.06567} \BibitemShut {NoStop}%
\bibitem [{\citenamefont {Escribano}\ \emph {et~al.}(2020)\citenamefont
  {Escribano}, \citenamefont {Prada}, \citenamefont {Oreg},\ and\ \citenamefont
  {Yeyati}}]{Escribano2020}%
  \BibitemOpen
  \bibfield  {author} {\bibinfo {author} {\bibfnamefont {S.~D.}\ \bibnamefont
  {Escribano}}, \bibinfo {author} {\bibfnamefont {E.}~\bibnamefont {Prada}},
  \bibinfo {author} {\bibfnamefont {Y.}~\bibnamefont {Oreg}}, \ and\ \bibinfo
  {author} {\bibfnamefont {A.~L.}\ \bibnamefont {Yeyati}},\ }\href
  {http://arxiv.org/abs/2011.06566} {\  (\bibinfo {year} {2020})},\ \Eprint
  {http://arxiv.org/abs/2011.06566} {arXiv:2011.06566} \BibitemShut {NoStop}%
\bibitem [{\citenamefont {Maiani}\ \emph {et~al.}(2021)\citenamefont {Maiani},
  \citenamefont {Seoane~Souto}, \citenamefont {Leijnse},\ and\ \citenamefont
  {Flensberg}}]{Maiani2020}%
  \BibitemOpen
  \bibfield  {author} {\bibinfo {author} {\bibfnamefont {A.}~\bibnamefont
  {Maiani}}, \bibinfo {author} {\bibfnamefont {R.}~\bibnamefont
  {Seoane~Souto}}, \bibinfo {author} {\bibfnamefont {M.}~\bibnamefont
  {Leijnse}}, \ and\ \bibinfo {author} {\bibfnamefont {K.}~\bibnamefont
  {Flensberg}},\ }\href {\doibase 10.1103/PhysRevB.103.104508} {\bibfield
  {journal} {\bibinfo  {journal} {Phys. Rev. B}\ }\textbf {\bibinfo {volume}
  {103}},\ \bibinfo {pages} {104508} (\bibinfo {year} {2021})}\BibitemShut
  {NoStop}%
\bibitem [{\citenamefont {P{\"{o}}yh{\"{o}}nen}\ \emph
  {et~al.}(2020)\citenamefont {P{\"{o}}yh{\"{o}}nen}, \citenamefont {Varjas},
  \citenamefont {Wimmer},\ and\ \citenamefont {Akhmerov}}]{Poyhonen2020}%
  \BibitemOpen
  \bibfield  {author} {\bibinfo {author} {\bibfnamefont {K.}~\bibnamefont
  {P{\"{o}}yh{\"{o}}nen}}, \bibinfo {author} {\bibfnamefont {D.}~\bibnamefont
  {Varjas}}, \bibinfo {author} {\bibfnamefont {M.}~\bibnamefont {Wimmer}}, \
  and\ \bibinfo {author} {\bibfnamefont {A.~R.}\ \bibnamefont {Akhmerov}},\
  }\href {http://arxiv.org/abs/2011.08263} {\  (\bibinfo {year} {2020})},\
  \Eprint {http://arxiv.org/abs/2011.08263} {arXiv:2011.08263} \BibitemShut
  {NoStop}%
\bibitem [{\citenamefont {Langbehn}\ \emph {et~al.}(2020)\citenamefont
  {Langbehn}, \citenamefont {Gonzalez}, \citenamefont {Brouwer},\ and\
  \citenamefont {von Oppen}}]{Langbehn2020}%
  \BibitemOpen
  \bibfield  {author} {\bibinfo {author} {\bibfnamefont {J.}~\bibnamefont
  {Langbehn}}, \bibinfo {author} {\bibfnamefont {S.~A.}\ \bibnamefont
  {Gonzalez}}, \bibinfo {author} {\bibfnamefont {P.~W.}\ \bibnamefont
  {Brouwer}}, \ and\ \bibinfo {author} {\bibfnamefont {F.}~\bibnamefont {von
  Oppen}},\ }\href {http://arxiv.org/abs/2012.00055} {\  (\bibinfo {year}
  {2020})},\ \Eprint {http://arxiv.org/abs/2012.00055} {arXiv:2012.00055}
  \BibitemShut {NoStop}%
\bibitem [{\citenamefont {Sau}\ \emph {et~al.}(2010)\citenamefont {Sau},
  \citenamefont {Lutchyn}, \citenamefont {Tewari},\ and\ \citenamefont {{Das
  Sarma}}}]{Sau2010}%
  \BibitemOpen
  \bibfield  {author} {\bibinfo {author} {\bibfnamefont {J.~D.}\ \bibnamefont
  {Sau}}, \bibinfo {author} {\bibfnamefont {R.~M.}\ \bibnamefont {Lutchyn}},
  \bibinfo {author} {\bibfnamefont {S.}~\bibnamefont {Tewari}}, \ and\ \bibinfo
  {author} {\bibfnamefont {S.}~\bibnamefont {{Das Sarma}}},\ }\href {\doibase
  10.1103/PhysRevLett.104.040502} {\bibfield  {journal} {\bibinfo  {journal}
  {Phys. Rev. Lett.}\ }\textbf {\bibinfo {volume} {104}},\ \bibinfo {pages}
  {040502} (\bibinfo {year} {2010})},\ \Eprint {http://arxiv.org/abs/0907.2239}
  {arXiv:0907.2239} \BibitemShut {NoStop}%
\bibitem [{\citenamefont {Tokuyasu}\ \emph {et~al.}(1988)\citenamefont
  {Tokuyasu}, \citenamefont {Sauls},\ and\ \citenamefont
  {Rainer}}]{Tokuyasu1988}%
  \BibitemOpen
  \bibfield  {author} {\bibinfo {author} {\bibfnamefont {T.}~\bibnamefont
  {Tokuyasu}}, \bibinfo {author} {\bibfnamefont {J.~A.}\ \bibnamefont {Sauls}},
  \ and\ \bibinfo {author} {\bibfnamefont {D.}~\bibnamefont {Rainer}},\ }\href
  {\doibase 10.1103/PhysRevB.38.8823} {\bibfield  {journal} {\bibinfo
  {journal} {Phys. Rev. B}\ }\textbf {\bibinfo {volume} {38}},\ \bibinfo
  {pages} {8823} (\bibinfo {year} {1988})}\BibitemShut {NoStop}%
\bibitem [{\citenamefont {Sau}\ \emph {et~al.}(2012)\citenamefont {Sau},
  \citenamefont {Tewari},\ and\ \citenamefont {{Das Sarma}}}]{Sau2012}%
  \BibitemOpen
  \bibfield  {author} {\bibinfo {author} {\bibfnamefont {J.~D.}\ \bibnamefont
  {Sau}}, \bibinfo {author} {\bibfnamefont {S.}~\bibnamefont {Tewari}}, \ and\
  \bibinfo {author} {\bibfnamefont {S.}~\bibnamefont {{Das Sarma}}},\ }\href
  {\doibase 10.1103/PhysRevB.85.064512} {\bibfield  {journal} {\bibinfo
  {journal} {Phys. Rev. B - Condens. Matter Mater. Phys.}\ }\textbf {\bibinfo
  {volume} {85}},\ \bibinfo {pages} {064512} (\bibinfo {year} {2012})},\
  \Eprint {http://arxiv.org/abs/1111.2054} {arXiv:1111.2054} \BibitemShut
  {NoStop}%
\bibitem [{\citenamefont {Stanev}\ and\ \citenamefont
  {Galitski}(2014)}]{Stanev2014}%
  \BibitemOpen
  \bibfield  {author} {\bibinfo {author} {\bibfnamefont {V.}~\bibnamefont
  {Stanev}}\ and\ \bibinfo {author} {\bibfnamefont {V.}~\bibnamefont
  {Galitski}},\ }\href {\doibase 10.1103/PhysRevB.89.174521} {\bibfield
  {journal} {\bibinfo  {journal} {Phys. Rev. B}\ }\textbf {\bibinfo {volume}
  {89}},\ \bibinfo {pages} {174521} (\bibinfo {year} {2014})},\ \Eprint
  {http://arxiv.org/abs/1402.5337} {arXiv:1402.5337} \BibitemShut {NoStop}%
\bibitem [{\citenamefont {Hui}\ \emph {et~al.}(2014)\citenamefont {Hui},
  \citenamefont {Sau},\ and\ \citenamefont {{Das Sarma}}}]{Hui2014}%
  \BibitemOpen
  \bibfield  {author} {\bibinfo {author} {\bibfnamefont {H.~Y.}\ \bibnamefont
  {Hui}}, \bibinfo {author} {\bibfnamefont {J.~D.}\ \bibnamefont {Sau}}, \ and\
  \bibinfo {author} {\bibfnamefont {S.}~\bibnamefont {{Das Sarma}}},\ }\href
  {\doibase 10.1103/PhysRevB.90.064516} {\bibfield  {journal} {\bibinfo
  {journal} {Phys. Rev. B - Condens. Matter Mater. Phys.}\ }\textbf {\bibinfo
  {volume} {90}},\ \bibinfo {pages} {064516} (\bibinfo {year} {2014})},\
  \Eprint {http://arxiv.org/abs/1406.4853} {arXiv:1406.4853} \BibitemShut
  {NoStop}%
\bibitem [{\citenamefont {Neven}\ \emph {et~al.}(2013)\citenamefont {Neven},
  \citenamefont {Bagrets},\ and\ \citenamefont {Altland}}]{Neven2013}%
  \BibitemOpen
  \bibfield  {author} {\bibinfo {author} {\bibfnamefont {P.}~\bibnamefont
  {Neven}}, \bibinfo {author} {\bibfnamefont {D.}~\bibnamefont {Bagrets}}, \
  and\ \bibinfo {author} {\bibfnamefont {A.}~\bibnamefont {Altland}},\ }\href
  {\doibase 10.1088/1367-2630/15/5/055019} {\bibfield  {journal} {\bibinfo
  {journal} {New J. Phys.}\ }\textbf {\bibinfo {volume} {15}},\ \bibinfo
  {pages} {055019} (\bibinfo {year} {2013})},\ \Eprint
  {http://arxiv.org/abs/1302.0747} {arXiv:1302.0747} \BibitemShut {NoStop}%
\bibitem [{\citenamefont {Lu}\ \emph {et~al.}(2020)\citenamefont {Lu},
  \citenamefont {Virtanen},\ and\ \citenamefont {Heikkil\"a}}]{Lu2020}%
  \BibitemOpen
  \bibfield  {author} {\bibinfo {author} {\bibfnamefont {Y.}~\bibnamefont
  {Lu}}, \bibinfo {author} {\bibfnamefont {P.}~\bibnamefont {Virtanen}}, \ and\
  \bibinfo {author} {\bibfnamefont {T.~T.}\ \bibnamefont {Heikkil\"a}},\ }\href
  {\doibase 10.1103/PhysRevB.102.224510} {\bibfield  {journal} {\bibinfo
  {journal} {Phys. Rev. B}\ }\textbf {\bibinfo {volume} {102}},\ \bibinfo
  {pages} {224510} (\bibinfo {year} {2020})}\BibitemShut {NoStop}%
\bibitem [{\citenamefont {Usadel}(1970)}]{Usadel1970}%
  \BibitemOpen
  \bibfield  {author} {\bibinfo {author} {\bibfnamefont {K.~D.}\ \bibnamefont
  {Usadel}},\ }\href {\doibase 10.1103/PhysRevLett.25.507} {\bibfield
  {journal} {\bibinfo  {journal} {Phys. Rev. Lett.}\ }\textbf {\bibinfo
  {volume} {25}},\ \bibinfo {pages} {507} (\bibinfo {year} {1970})}\BibitemShut
  {NoStop}%
\bibitem [{\citenamefont {Ivanov}\ and\ \citenamefont
  {Fominov}(2006)}]{Ivanov2006}%
  \BibitemOpen
  \bibfield  {author} {\bibinfo {author} {\bibfnamefont {D.~A.}\ \bibnamefont
  {Ivanov}}\ and\ \bibinfo {author} {\bibfnamefont {Y.~V.}\ \bibnamefont
  {Fominov}},\ }\href {\doibase 10.1103/PhysRevB.73.214524} {\bibfield
  {journal} {\bibinfo  {journal} {Phys. Rev. B - Condens. Matter Mater. Phys.}\
  }\textbf {\bibinfo {volume} {73}},\ \bibinfo {pages} {214524} (\bibinfo
  {year} {2006})},\ \Eprint {http://arxiv.org/abs/0511299} {arXiv:0511299
  [cond-mat]} \BibitemShut {NoStop}%
\bibitem [{\citenamefont {Bergeret}\ \emph {et~al.}(2005)\citenamefont
  {Bergeret}, \citenamefont {Volkov},\ and\ \citenamefont
  {Efetov}}]{Bergeret2005}%
  \BibitemOpen
  \bibfield  {author} {\bibinfo {author} {\bibfnamefont {F.~S.}\ \bibnamefont
  {Bergeret}}, \bibinfo {author} {\bibfnamefont {A.~F.}\ \bibnamefont
  {Volkov}}, \ and\ \bibinfo {author} {\bibfnamefont {K.~B.}\ \bibnamefont
  {Efetov}},\ }\href {\doibase 10.1103/RevModPhys.77.1321} {\bibfield
  {journal} {\bibinfo  {journal} {Rev. Mod. Phys.}\ }\textbf {\bibinfo {volume}
  {77}},\ \bibinfo {pages} {1321} (\bibinfo {year} {2005})},\ \Eprint
  {http://arxiv.org/abs/0506047} {arXiv:0506047 [cond-mat]} \BibitemShut
  {NoStop}%
\bibitem [{\citenamefont {Gall}(2016)}]{Gall2016}%
  \BibitemOpen
  \bibfield  {author} {\bibinfo {author} {\bibfnamefont {D.}~\bibnamefont
  {Gall}},\ }\href {\doibase 10.1063/1.4942216} {\bibfield  {journal} {\bibinfo
   {journal} {J. Appl. Phys.}\ }\textbf {\bibinfo {volume} {119}},\ \bibinfo
  {pages} {085101} (\bibinfo {year} {2016})}\BibitemShut {NoStop}%
\bibitem [{\citenamefont {Romijn}\ \emph {et~al.}(1982)\citenamefont {Romijn},
  \citenamefont {Klapwijk}, \citenamefont {Renne},\ and\ \citenamefont
  {Mooij}}]{Romijn1982}%
  \BibitemOpen
  \bibfield  {author} {\bibinfo {author} {\bibfnamefont {J.}~\bibnamefont
  {Romijn}}, \bibinfo {author} {\bibfnamefont {T.~M.}\ \bibnamefont
  {Klapwijk}}, \bibinfo {author} {\bibfnamefont {M.~J.}\ \bibnamefont {Renne}},
  \ and\ \bibinfo {author} {\bibfnamefont {J.~E.}\ \bibnamefont {Mooij}},\
  }\href {\doibase 10.1103/PhysRevB.26.3648} {\bibfield  {journal} {\bibinfo
  {journal} {Phys. Rev. B}\ }\textbf {\bibinfo {volume} {26}},\ \bibinfo
  {pages} {3648} (\bibinfo {year} {1982})}\BibitemShut {NoStop}%
\bibitem [{\citenamefont {Gorkov}(1958)}]{Gorkov1958}%
  \BibitemOpen
  \bibfield  {author} {\bibinfo {author} {\bibfnamefont {L.}~\bibnamefont
  {Gorkov}},\ }\href {http://www.jetp.ac.ru/cgi-bin/dn/e_007_03_0505.pdf}
  {\bibfield  {journal} {\bibinfo  {journal} {Sov. Phys. JETP}\ }\textbf
  {\bibinfo {volume} {34}},\ \bibinfo {pages} {505} (\bibinfo {year}
  {1958})}\BibitemShut {NoStop}%
\bibitem [{\citenamefont {Eilenberger}(1968)}]{Eilenberger1968}%
  \BibitemOpen
  \bibfield  {author} {\bibinfo {author} {\bibfnamefont {G.}~\bibnamefont
  {Eilenberger}},\ }\href {\doibase 10.1007/BF01379803} {\bibfield  {journal}
  {\bibinfo  {journal} {Zeitschrift f{\"{u}}r Phys.}\ }\textbf {\bibinfo
  {volume} {214}},\ \bibinfo {pages} {195} (\bibinfo {year}
  {1968})}\BibitemShut {NoStop}%
\bibitem [{\citenamefont {Larkin}\ and\ \citenamefont
  {Ovchinnikov}(1969)}]{Larkin1969}%
  \BibitemOpen
  \bibfield  {author} {\bibinfo {author} {\bibfnamefont {A.}~\bibnamefont
  {Larkin}}\ and\ \bibinfo {author} {\bibfnamefont {Y.}~\bibnamefont
  {Ovchinnikov}},\ }\href
  {http://www.jetp.ac.ru/cgi-bin/e/index/e/28/6/p1200?a=list} {\bibfield
  {journal} {\bibinfo  {journal} {Sov Phys JETP}\ }\textbf {\bibinfo {volume}
  {28}},\ \bibinfo {pages} {1200} (\bibinfo {year} {1969})}\BibitemShut
  {NoStop}%
\bibitem [{\citenamefont {Alexander}\ \emph {et~al.}(1985)\citenamefont
  {Alexander}, \citenamefont {Orlando}, \citenamefont {Rainer},\ and\
  \citenamefont {Tedrow}}]{Alexander1985}%
  \BibitemOpen
  \bibfield  {author} {\bibinfo {author} {\bibfnamefont {J.~A.~X.}\
  \bibnamefont {Alexander}}, \bibinfo {author} {\bibfnamefont {T.~P.}\
  \bibnamefont {Orlando}}, \bibinfo {author} {\bibfnamefont {D.}~\bibnamefont
  {Rainer}}, \ and\ \bibinfo {author} {\bibfnamefont {P.~M.}\ \bibnamefont
  {Tedrow}},\ }\href {\doibase 10.1103/PhysRevB.31.5811} {\bibfield  {journal}
  {\bibinfo  {journal} {Phys. Rev. B}\ }\textbf {\bibinfo {volume} {31}},\
  \bibinfo {pages} {5811} (\bibinfo {year} {1985})}\BibitemShut {NoStop}%
\bibitem [{\citenamefont {Demler}\ \emph {et~al.}(1997)\citenamefont {Demler},
  \citenamefont {Arnold},\ and\ \citenamefont {Beasley}}]{Demler1997}%
  \BibitemOpen
  \bibfield  {author} {\bibinfo {author} {\bibfnamefont {E.~A.}\ \bibnamefont
  {Demler}}, \bibinfo {author} {\bibfnamefont {G.~B.}\ \bibnamefont {Arnold}},
  \ and\ \bibinfo {author} {\bibfnamefont {M.~R.}\ \bibnamefont {Beasley}},\
  }\href {\doibase 10.1103/PhysRevB.55.15174} {\bibfield  {journal} {\bibinfo
  {journal} {Phys. Rev. B}\ }\textbf {\bibinfo {volume} {55}},\ \bibinfo
  {pages} {15174} (\bibinfo {year} {1997})}\BibitemShut {NoStop}%
\bibitem [{\citenamefont {Aikebaier}\ \emph {et~al.}(2019)\citenamefont
  {Aikebaier}, \citenamefont {Virtanen},\ and\ \citenamefont
  {Heikkil{\"{a}}}}]{Aikebaier2019}%
  \BibitemOpen
  \bibfield  {author} {\bibinfo {author} {\bibfnamefont {F.}~\bibnamefont
  {Aikebaier}}, \bibinfo {author} {\bibfnamefont {P.}~\bibnamefont {Virtanen}},
  \ and\ \bibinfo {author} {\bibfnamefont {T.}~\bibnamefont {Heikkil{\"{a}}}},\
  }\href {\doibase 10.1103/PhysRevB.99.104504} {\bibfield  {journal} {\bibinfo
  {journal} {Phys. Rev. B}\ }\textbf {\bibinfo {volume} {99}},\ \bibinfo
  {pages} {104504} (\bibinfo {year} {2019})},\ \Eprint
  {http://arxiv.org/abs/1812.08410} {arXiv:1812.08410} \BibitemShut {NoStop}%
\bibitem [{\citenamefont {Kiendl}\ \emph {et~al.}(2019)\citenamefont {Kiendl},
  \citenamefont {{Von Oppen}},\ and\ \citenamefont {Brouwer}}]{Kiendl2019}%
  \BibitemOpen
  \bibfield  {author} {\bibinfo {author} {\bibfnamefont {T.}~\bibnamefont
  {Kiendl}}, \bibinfo {author} {\bibfnamefont {F.}~\bibnamefont {{Von Oppen}}},
  \ and\ \bibinfo {author} {\bibfnamefont {P.~W.}\ \bibnamefont {Brouwer}},\
  }\href {\doibase 10.1103/PhysRevB.100.035426} {\bibfield  {journal} {\bibinfo
   {journal} {Phys. Rev. B}\ }\textbf {\bibinfo {volume} {100}},\ \bibinfo
  {pages} {035426} (\bibinfo {year} {2019})},\ \Eprint
  {http://arxiv.org/abs/1902.09798} {arXiv:1902.09798} \BibitemShut {NoStop}%
\bibitem [{\citenamefont {Liu}\ \emph {et~al.}(2018)\citenamefont {Liu},
  \citenamefont {Rossi},\ and\ \citenamefont {Lutchyn}}]{Liu2018}%
  \BibitemOpen
  \bibfield  {author} {\bibinfo {author} {\bibfnamefont {D.~E.}\ \bibnamefont
  {Liu}}, \bibinfo {author} {\bibfnamefont {E.}~\bibnamefont {Rossi}}, \ and\
  \bibinfo {author} {\bibfnamefont {R.~M.}\ \bibnamefont {Lutchyn}},\ }\href
  {\doibase 10.1103/PhysRevB.97.161408} {\bibfield  {journal} {\bibinfo
  {journal} {Phys. Rev. B}\ }\textbf {\bibinfo {volume} {97}},\ \bibinfo
  {pages} {161408} (\bibinfo {year} {2018})},\ \Eprint
  {http://arxiv.org/abs/1711.04056} {arXiv:1711.04056} \BibitemShut {NoStop}%
\bibitem [{\citenamefont {Cottet}\ \emph {et~al.}(2009)\citenamefont {Cottet},
  \citenamefont {Huertas-Hernando}, \citenamefont {Belzig},\ and\ \citenamefont
  {Nazarov}}]{Cottet2009}%
  \BibitemOpen
  \bibfield  {author} {\bibinfo {author} {\bibfnamefont {A.}~\bibnamefont
  {Cottet}}, \bibinfo {author} {\bibfnamefont {D.}~\bibnamefont
  {Huertas-Hernando}}, \bibinfo {author} {\bibfnamefont {W.}~\bibnamefont
  {Belzig}}, \ and\ \bibinfo {author} {\bibfnamefont {Y.~V.}\ \bibnamefont
  {Nazarov}},\ }\href {\doibase 10.1103/PhysRevB.80.184511} {\bibfield
  {journal} {\bibinfo  {journal} {Phys. Rev. B}\ }\textbf {\bibinfo {volume}
  {80}},\ \bibinfo {pages} {184511} (\bibinfo {year} {2009})}\BibitemShut
  {NoStop}%
\bibitem [{Note1()}]{Note1}%
  \BibitemOpen
  \bibinfo {note} {SC-MI bilayers with superconductors of thickness comparable
  or greater than the coherence length have been recently studied in Ref.~\cite
  {Hijano2021}}\BibitemShut {NoStop}%
\bibitem [{\citenamefont {Stanescu}\ \emph {et~al.}(2011)\citenamefont
  {Stanescu}, \citenamefont {Lutchyn},\ and\ \citenamefont {{Das
  Sarma}}}]{Stanescu2011}%
  \BibitemOpen
  \bibfield  {author} {\bibinfo {author} {\bibfnamefont {T.~D.}\ \bibnamefont
  {Stanescu}}, \bibinfo {author} {\bibfnamefont {R.~M.}\ \bibnamefont
  {Lutchyn}}, \ and\ \bibinfo {author} {\bibfnamefont {S.}~\bibnamefont {{Das
  Sarma}}},\ }\href {\doibase 10.1103/PhysRevB.84.144522} {\bibfield  {journal}
  {\bibinfo  {journal} {Phys. Rev. B - Condens. Matter Mater. Phys.}\ }\textbf
  {\bibinfo {volume} {84}},\ \bibinfo {pages} {144522} (\bibinfo {year}
  {2011})},\ \Eprint {http://arxiv.org/abs/1106.3078} {arXiv:1106.3078}
  \BibitemShut {NoStop}%
\bibitem [{\citenamefont {Potter}\ and\ \citenamefont
  {Lee}(2011)}]{Potter2011}%
  \BibitemOpen
  \bibfield  {author} {\bibinfo {author} {\bibfnamefont {A.~C.}\ \bibnamefont
  {Potter}}\ and\ \bibinfo {author} {\bibfnamefont {P.~A.}\ \bibnamefont
  {Lee}},\ }\href {\doibase 10.1103/PhysRevB.83.184520} {\bibfield  {journal}
  {\bibinfo  {journal} {Phys. Rev. B - Condens. Matter Mater. Phys.}\ }\textbf
  {\bibinfo {volume} {83}},\ \bibinfo {pages} {184520} (\bibinfo {year}
  {2011})},\ \Eprint {http://arxiv.org/abs/1103.2129} {arXiv:1103.2129}
  \BibitemShut {NoStop}%
\bibitem [{Note2()}]{Note2}%
  \BibitemOpen
  \bibinfo {note} {Spin-dependent tunneling between the SC and SM has been
  recently considered in Ref.~\cite {Maiani2020}}\BibitemShut {NoStop}%
\bibitem [{\citenamefont {Sarma}(1963)}]{Sarma1963}%
  \BibitemOpen
  \bibfield  {author} {\bibinfo {author} {\bibfnamefont {G.}~\bibnamefont
  {Sarma}},\ }\href {\doibase 10.1016/0022-3697(63)90007-6} {\bibfield
  {journal} {\bibinfo  {journal} {J. Phys. Chem. Solids}\ }\textbf {\bibinfo
  {volume} {24}},\ \bibinfo {pages} {1029} (\bibinfo {year}
  {1963})}\BibitemShut {NoStop}%
\bibitem [{\citenamefont {Bruno}\ and\ \citenamefont
  {Schwartz}(1973)}]{Bruno1973}%
  \BibitemOpen
  \bibfield  {author} {\bibinfo {author} {\bibfnamefont {R.~C.}\ \bibnamefont
  {Bruno}}\ and\ \bibinfo {author} {\bibfnamefont {B.~B.}\ \bibnamefont
  {Schwartz}},\ }\href {\doibase 10.1103/PhysRevB.8.3161} {\bibfield  {journal}
  {\bibinfo  {journal} {Phys. Rev. B}\ }\textbf {\bibinfo {volume} {8}},\
  \bibinfo {pages} {3161} (\bibinfo {year} {1973})}\BibitemShut {NoStop}%
\bibitem [{\citenamefont {Bergeret}\ \emph {et~al.}(2018)\citenamefont
  {Bergeret}, \citenamefont {Silaev}, \citenamefont {Virtanen},\ and\
  \citenamefont {Heikkil{\"{a}}}}]{Bergeret2018}%
  \BibitemOpen
  \bibfield  {author} {\bibinfo {author} {\bibfnamefont {F.~S.}\ \bibnamefont
  {Bergeret}}, \bibinfo {author} {\bibfnamefont {M.}~\bibnamefont {Silaev}},
  \bibinfo {author} {\bibfnamefont {P.}~\bibnamefont {Virtanen}}, \ and\
  \bibinfo {author} {\bibfnamefont {T.~T.}\ \bibnamefont {Heikkil{\"{a}}}},\
  }\href {\doibase 10.1103/RevModPhys.90.041001} {\bibfield  {journal}
  {\bibinfo  {journal} {Rev. Mod. Phys.}\ }\textbf {\bibinfo {volume} {90}},\
  \bibinfo {pages} {041001} (\bibinfo {year} {2018})}\BibitemShut {NoStop}%
\bibitem [{\citenamefont {Heikkil{\"{a}}}\ \emph {et~al.}(2019)\citenamefont
  {Heikkil{\"{a}}}, \citenamefont {Silaev}, \citenamefont {Virtanen},\ and\
  \citenamefont {Bergeret}}]{Heikkila2019}%
  \BibitemOpen
  \bibfield  {author} {\bibinfo {author} {\bibfnamefont {T.~T.}\ \bibnamefont
  {Heikkil{\"{a}}}}, \bibinfo {author} {\bibfnamefont {M.}~\bibnamefont
  {Silaev}}, \bibinfo {author} {\bibfnamefont {P.}~\bibnamefont {Virtanen}}, \
  and\ \bibinfo {author} {\bibfnamefont {F.~S.}\ \bibnamefont {Bergeret}},\
  }\href {\doibase 10.1016/j.progsurf.2019.100540} {\bibfield  {journal}
  {\bibinfo  {journal} {Progress in Surface Science}\ }\textbf {\bibinfo
  {volume} {94}},\ \bibinfo {pages} {100540} (\bibinfo {year}
  {2019})}\BibitemShut {NoStop}%
\bibitem [{\citenamefont {Chandrasekhar}(1962)}]{Chandrasekhar1962}%
  \BibitemOpen
  \bibfield  {author} {\bibinfo {author} {\bibfnamefont {B.~S.}\ \bibnamefont
  {Chandrasekhar}},\ }\href {\doibase 10.1063/1.1777362} {\bibfield  {journal}
  {\bibinfo  {journal} {Appl. Phys. Lett.}\ }\textbf {\bibinfo {volume} {1}},\
  \bibinfo {pages} {7} (\bibinfo {year} {1962})}\BibitemShut {NoStop}%
\bibitem [{\citenamefont {Clogston}(1962)}]{Clogston1962}%
  \BibitemOpen
  \bibfield  {author} {\bibinfo {author} {\bibfnamefont {A.~M.}\ \bibnamefont
  {Clogston}},\ }\href {\doibase 10.1103/PhysRevLett.9.266} {\bibfield
  {journal} {\bibinfo  {journal} {Phys. Rev. Lett.}\ }\textbf {\bibinfo
  {volume} {9}},\ \bibinfo {pages} {266} (\bibinfo {year} {1962})}\BibitemShut
  {NoStop}%
\bibitem [{\citenamefont {Tinkham}(2004)}]{Tinkham2004}%
  \BibitemOpen
  \bibfield  {author} {\bibinfo {author} {\bibfnamefont {M.}~\bibnamefont
  {Tinkham}},\ }\href {https://store.doverpublications.com/0486435032.html}
  {\emph {\bibinfo {title} {Introd. to Supercond. 2nd Ed.}}},\ \bibinfo
  {edition} {2nd}\ ed.\ (\bibinfo  {publisher} {Dover},\ \bibinfo {year}
  {2004})\BibitemShut {NoStop}%
\bibitem [{\citenamefont {Kanne}\ \emph {et~al.}(2020)\citenamefont {Kanne},
  \citenamefont {Marnauza}, \citenamefont {Olsteins}, \citenamefont {Carrad},
  \citenamefont {Sestoft}, \citenamefont {de~Bruijckere}, \citenamefont {Zeng},
  \citenamefont {Johnson}, \citenamefont {Olsson}, \citenamefont
  {Grove-Rasmussen},\ and\ \citenamefont {Nyg{\aa}rd}}]{Kanne2020}%
  \BibitemOpen
  \bibfield  {author} {\bibinfo {author} {\bibfnamefont {T.}~\bibnamefont
  {Kanne}}, \bibinfo {author} {\bibfnamefont {M.}~\bibnamefont {Marnauza}},
  \bibinfo {author} {\bibfnamefont {D.}~\bibnamefont {Olsteins}}, \bibinfo
  {author} {\bibfnamefont {D.~J.}\ \bibnamefont {Carrad}}, \bibinfo {author}
  {\bibfnamefont {J.~E.}\ \bibnamefont {Sestoft}}, \bibinfo {author}
  {\bibfnamefont {J.}~\bibnamefont {de~Bruijckere}}, \bibinfo {author}
  {\bibfnamefont {L.}~\bibnamefont {Zeng}}, \bibinfo {author} {\bibfnamefont
  {E.}~\bibnamefont {Johnson}}, \bibinfo {author} {\bibfnamefont
  {E.}~\bibnamefont {Olsson}}, \bibinfo {author} {\bibfnamefont
  {K.}~\bibnamefont {Grove-Rasmussen}}, \ and\ \bibinfo {author} {\bibfnamefont
  {J.}~\bibnamefont {Nyg{\aa}rd}},\ }\href {http://arxiv.org/abs/2002.11641}
  {\bibfield  {journal} {\bibinfo  {journal} {arXiv}\ } (\bibinfo {year}
  {2020})},\ \Eprint {http://arxiv.org/abs/2002.11641} {arXiv:2002.11641}
  \BibitemShut {NoStop}%
\end{thebibliography}%

\onecolumngrid

\appendix

\section{Analytical expressions for the proximity-induced terms of the nanowire Green's function in the absence of spin-orbit and magnetic scattering in the SC}
\label{sec:Zeeman_analyt}

In the absence of magnetic and spin-orbit scattering $\Gamma_{so}=\Gamma_{sf}=0$ pair potential in the superconductor is constant as a function of Zeeman energy $\Delta=\Delta_{00}$ up to the Clogston limit, and analytical solution to the Usadel equations \eqref{eq:Usad_alg1}-\eqref{eq:Usad_alg2} can be obtained\cite{Aikebaier2019}:
\begin{align}
	\tan\theta &=\frac{\sqrt{4\omega_n^2\Delta^2+\left[(V_Z^{SC})^2+\omega_n^2-\Delta^2\right]^2}-(V_Z^{SC})^2-\omega_n^2+\Delta^2}{2\omega_n\Delta} \label{eq:Usad_analyt1} \\
	\cosh\phi &=\frac{\omega_n+\Delta\tan\theta}{\sqrt{\omega_n^2+\left[\Delta^2-(V_Z^{SC})^2\right]\tan^2\theta+2\omega_n\Delta\tan\theta}} \label{eq:Usad_analyt2} \\
	\sinh\phi &=\frac{V_Z^{SC}\tan\theta}{\sqrt{\omega_n^2+\left[\Delta^2-(V_Z^{SC})^2\right]\tan^2\theta+2\omega_n\Delta\tan\theta}} \label{eq:Usad_analyt3}
\end{align}
Performing analytical continuation of the Matsubara frequencies $\omega_n\to -i\omega$, we obtain the proximity induced terms of the Green's function \eqref{eq:GF_NW_final}
\begin{align}
	&i\cos\theta\cosh\phi=-\frac{-\omega^2+S}{2\omega\sqrt{1-\frac{(\omega^2+S)^2}{4\Delta^2\omega^2}}\sqrt{S-\frac{(-h^2+\Delta^2)(\omega^2+S)^2}{4\Delta^2\omega^2}}} \\
	&\sin\theta\sinh\phi=-\frac{h(\omega^2+S)^2}{4\Delta^2\omega^2\sqrt{1-\frac{(\omega^2+S)^2}{4\Delta^2\omega^2}}\sqrt{S-\frac{(-h^2+\Delta^2)(\omega^2+S)^2}{4\Delta^2\omega^2}}}    \\
	&\sin\theta\cosh\phi=-\frac{(-\omega^2+S)(\omega^2+S)}{4\Delta\omega^2\sqrt{1-\frac{(\omega^2+S)^2}{4\Delta^2\omega^2}}\sqrt{S-\frac{(-h^2+\Delta^2)(\omega^2+S)^2}{4\Delta^2\omega^2}}}    \\
	&i\cos\theta\sinh\phi=-\frac{h(\omega^2+S)^2}{2\Delta\omega\sqrt{1-\frac{(\omega^2+S)^2}{4\Delta^2\omega^2}}\sqrt{S-\frac{(-h^2+\Delta^2)(\omega^2+S)^2}{4\Delta^2\omega^2}}}
\end{align}
where for notational purposes we denoted $S=-(V_Z^{SC})^2+\Delta^2+\sqrt{-4\Delta^2\omega^2+\left(\omega^2-(V_Z^{SC})^2+\Delta^2\right)^2}$.

\end{document}